\documentclass[apjl,usenatbib,usegraphicx]{emulateapj-rtx4}
\usepackage{graphicx,amssymb}

\newcommand{\lamost}{{\sc lamost}}
\newcommand{\kasc}{{\sc kasc}}
\newcommand{\kepler}{{\it Kepler}}
\newcommand{\degree}{$^{\circ}$}
\newcommand{\teff}{$T_{\rm eff}$}
\newcommand{\logg}{$\log g$}
\newcommand{\vsini}{$v \sin i$}
\newcommand{\mh}{$[M/H]$}
\newcommand{\feh}{$[Fe/H]$}
\newcommand{\ebv}{$E(B-V)$}
\newcommand{\pointing}{LK-field}
\newcommand{\pointings}{LK-fields}
\newcommand{\project}{LK-project}
\newcommand{\Kp}{$K\rm p$}

\newcommand{\snrr}{SNR$_r$}

\newcommand{\vrad}{$v_{\rm rad}$}

\newcommand{\rotfit}{{\sc rotfit}}
\newcommand{\mkclass}{{\sc mkclass}}
\newcommand{\ulyss}{{\sc ul}y{\sc ss}}
\newcommand{\cfi}{{\sc cfi}}
\newcommand{\lasp}{{\sc lasp}}
\newcommand{\kms}{km\,s$^{-1}$}

\shorttitle{\lamost\ observations in the \kepler\ field. Database of low-resolution spectra}
\shortauthors{De Cat et al.}

\begin{document}

\title{
\lamost\ observations in the \kepler\ field. Database of low-resolution spectra\footnote{Based on observations collected using the Large Sky Area Multi-Object Fiber Spectroscopic Telescope (\lamost) located at the Xinglong observatory, China.}
}

\author{
P. De Cat\altaffilmark{2}\email{Peter.DeCat@oma.be}, 
J.N. Fu\altaffilmark{1}, 
A.B. Ren\altaffilmark{1}, 
X.H. Yang\altaffilmark{1}, 
J.R. Shi\altaffilmark{3}, %10 nights 
A.L. Luo\altaffilmark{3}, 
M. Yang\altaffilmark{3}, %7 nights
J.L. Wang\altaffilmark{3}, %6 nights
H.T. Zhang\altaffilmark{3}, %3 nights
H.M. Shi\altaffilmark{3}, %1 night
W. Zhang\altaffilmark{3}, 
Subo Dong\altaffilmark{4}, 
G. Catanzaro\altaffilmark{5},
C.J. Corbally\altaffilmark{6}, 
A. Frasca\altaffilmark{5}, 
R.O. Gray\altaffilmark{7}, 
J. Molenda-\.Zakowicz\altaffilmark{8,9},
K. Uytterhoeven\altaffilmark{10,11}, 
M. Briquet\altaffilmark{12,13}, 
H. Bruntt\altaffilmark{14}, 
S. Frandsen\altaffilmark{14}, 
L. Kiss\altaffilmark{15},
D.W. Kurtz\altaffilmark{16}, 
M. Marconi\altaffilmark{17}, 
E. Niemczura\altaffilmark{8}, 
R.H. {\O}stensen\altaffilmark{18}, 
V. Ripepi\altaffilmark{17},
B. Smalley\altaffilmark{19}, 
J. Southworth\altaffilmark{19},
R. Szab\'o\altaffilmark{15}, 
J.H. Telting\altaffilmark{20},
C. Karoff\altaffilmark{14,21}, 
V. Silva Aguirre\altaffilmark{14}, 
Y. Wu\altaffilmark{3}, 
Y.H. Hou\altaffilmark{22}, 
G. Jin\altaffilmark{23}, 
X.L. Zhou\altaffilmark{24}
}

\altaffiltext{1}{Department of Astronomy, Beijing Normal University, 19 Avenue Xinjiekouwai, Beijing 100875, China}
\altaffiltext{2}{Royal observatory of Belgium, Ringlaan 3, B-1180 Brussel, Belgium}
\altaffiltext{3}{Key Lab for Optical Astronomy, National Astronomical Observatories, Chinese Academy of Sciences, Beijing 100012, China}
\altaffiltext{4}{Kavli Institute for Astronomy and Astrophysics, Peking University, Yi He Yuan Road 5, Hai Dian District, Beijing, 100871, China}
\altaffiltext{5}{INAF - Osservatorio Astrofisico di Catania, Via S. Sofia 78, 95123 Catania, Italy}
\altaffiltext{6}{Vatican Observatory Research Group, Steward Observatory, Tucson, AZ 85721-0065, USA}
\altaffiltext{7}{Department of Physics and Astronomy, Appalachian State University, Boone, NC 28608, USA}
\altaffiltext{8}{Astronomical Institute of the University of Wroc{\l}aw, ul. Kopernika 11, 51-622 Wroc{\l}aw, Poland}
\altaffiltext{9}{New Mexico State University, Department of Astronomy, P.O.Box 30001, MSC 4500, Las Cruces, NM 88003, USA}
\altaffiltext{10}{Instituto de Astrof\'{\i}sica de Canarias (IAC), E-38200 La Laguna, Tenerife, Spain}
\altaffiltext{11}{Universidad de La Laguna, Dept. Astrof\'{\i}sica, E-38206 La Laguna, Tenerife, Spain}
\altaffiltext{12}{Institut d'Astrophysique et de G\'eophysique, Universit\'e de Li\`ege, All\'ee du 6 Ao\^ut 19C, 4000 Li\`ege, Belgium}
\altaffiltext{13}{LESIA, Observatoire de Paris, PSL Research University, CNRS, Sorbonne Universit\'es, UPMC Univ. Paris 06, Univ. Paris Diderot, Sorbonne Paris Cit\'e}
\altaffiltext{14}{Stellar Astrophysics Center, Department of Physics and Astronomy, Aarhus University, Ny Munkegade 120, DK-8000 Aarhus C, Denmark}
\altaffiltext{15}{Konkoly Observatory, Research Center for Astronomy and Earth Sciences, Hungarian Academy of Sciences, Konkoly Thege Mikl\'os \'ut 15-17. H-1121 Budapest, Hungary}
\altaffiltext{16}{Jeremiah Horrocks Institute, University of Central Lancashire, Preston PR1\,2HE, UK}
\altaffiltext{17}{INAF - Osservatorio Astronomico di Capodimonte, via Moiariello 16, 80131 Napoli, Italy}
\altaffiltext{18}{Instituut voor Sterrenkunde, KU Leuven, Celestijnenlaan 200D, B-3001 Leuven, Belgium}
\altaffiltext{19}{Astrophysics Group, Keele University, Staffordshire, ST5 5BG, UK}
\altaffiltext{20}{Nordic Optical Telescope, Rambla Jos\'e Ana Fern\'andez P\'erez 7, 38711 Bre\~na Baja, Spain}
\altaffiltext{21}{Department of Geoscience, Aarhus University, H{\o}egh-Guldbergs Gade 2, 8000, Aarhus C, Denmark}
\altaffiltext{22}{Nanjing Institute of Astronomical Optics \& Technology, National Astronomical Observatories, Chinese Academy of Sciences, Nanjing 210042, China}
\altaffiltext{23}{University of Science and Technology of China, Hefei 230026, China}
\altaffiltext{24}{Key Laboratory of Space Astronomy and Technology, National Astronomical Observatories, Chinese Academy of Sciences, Beijing 100012, China}

\begin{abstract}
The nearly continuous light curves with micromagnitude precision provided by the space mission \kepler\ are revolutionising our view of pulsating stars. 
They have revealed a vast sea of low-amplitude pulsation modes that were undetectable from Earth. 
The long time base of \kepler\ light curves allows an accurate determination of frequencies and amplitudes of pulsation modes needed for in-depth asteroseismic modeling. 
However, for an asteroseismic study to be successful, the first estimates of stellar parameters need to be known and they can not be derived from the \kepler\ photometry itself. 
The \kepler\ Input Catalog (KIC) provides values for the effective temperature, the surface gravity and the metallicity, but not always with a sufficient accuracy. 
Moreover, information on the chemical composition and rotation rate is lacking. 
We are collecting low-resolution spectra for objects in the \kepler\ field of view with the Large Sky Area Multi-Object Fiber Spectroscopic Telescope (\lamost, Xinglong observatory, China). 
All of the requested fields have now been observed at least once.
In this paper we describe those observations and provide a database of use to the whole astronomical community. 
\end{abstract}


%%%%%%%%%%%%%%%%%%%%%%%%%%%%%%%%%%%%%%%%%%%%%%%%%%%%%%%%%%%%%%%%%%%%%%%%%%%%%%
\section{Introduction}
\label{sect:intro}
%%%%%%%%%%%%%%%%%%%%%%%%%%%%%%%%%%%%%%%%%%%%%%%%%%%%%%%%%%%%%%%%%%%%%%%%%%%%%%

The space mission \kepler\ was designed to detect Earth-like planets around solar-type stars by the transit method \citep{Koch2010ApJ...713L..79K}. 
It was launched on 2009 March 7 and started to collect ultraprecise photometry within a spectral bandpass from 400~nm to 850~nm for a fixed field of view (FoV) of 105 square degrees in the constellations Lyra and Cygnus on 2009 May 2. 
On 2013 May 11, a second of the four reaction wheels of the  \kepler\ spacecraft failed which prevents the telescope from precisely pointing towards the same FoV. 
Even though ultrahigh precision photometry can no longer be collected in the original \kepler\ FoV, the \kepler\ photometry that is now available for about 200\,000 stars is a pure goldmine for asteroseismic studies of all types of pulsating stars, as well as for many other science cases.

The success of asteroseismic studies has been shown to depend crucially on the availability of first estimates of basic stellar parameters, such as the effective temperature (\teff), surface gravity (\logg), metallicity (\mh), and the projected stellar equatorial rotation velocity (\vsini) \citep{Cunha2007A&ARv..14..217C,Michel2006CoAst.147...40M}. 
These parameters cannot always be derived in a direct way from the \kepler\ data as is the case for data with multi-color photometry or spectroscopy. 
Before the launch of the \kepler\ spacecraft, there was a large effort to derive the stellar parameters from Sloan photometry for potential \kepler\ targets. 
These are available in the \kepler\ Input Catalog (KIC; \citealt{Brown2011AJ....142..112B}). 
Unfortunately, KIC stellar parameters are not available for all the stars of interest and the precision of the \teff\ and \logg\ in KIC is generally too low for asteroseismic modeling, especially for hot and peculiar stars \citep{Molenda2010arXiv1005.0985M,McNamara2012AJ....143..101M}. 
Also, detailed information on the stellar chemical composition and rotation rate is lacking. 
Hence, to exploit the \kepler\ data best, additional ground-based {\it spectroscopic} data are required \citep{Uytterhoeven2010AN....331..993U,Uytterhoeven2010arXiv1003.6089U}. 

\begin{figure}
\begin{center}
 \resizebox{0.45\textwidth}{!}{\rotatebox{270}{\includegraphics{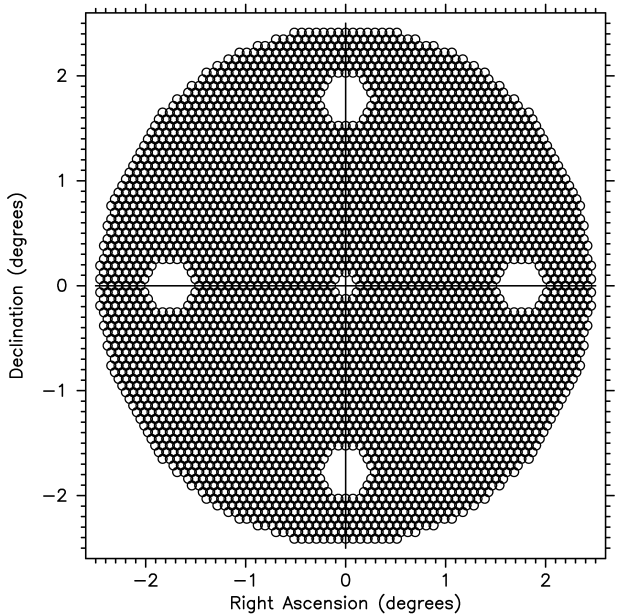}}}
 \end{center}
 \caption{\label{fig:fov} 
Configuration of the 4\,000 fiber positioning units covering the 5\degree\ \lamost\ focal plane. The fibers are homogeneously distributed on the focal plane except for the position of active optics wave front sensor (central hole) and 4 Guiding CCD cameras (off-center holes). }
\end{figure}

The Large Sky Area Multi-Object Fiber Spectroscopic Telescope (\lamost, also called the Guo Shou Jing Telescope) is a unique astronomical instrument located at the Xing\-long observatory (China). It combines a large aperture ($3.6-4.9$\,m) with a wide FoV (circular with diameters of 5\degree) \citep{Wang1996ApOpt..35.5155W}. The focal surface is covered with 4\,000 optical fibers connected to 16 multi-object optical spectrometers with 250 optical fibers each (\citealt{Xing1998SPIE.3352..839X}; Fig.\,\ref{fig:fov}). Each spectrometer has two CCD cameras: one for the blue arm (optimized for $370-590$\,nm) and one for the red arm (optimized for $570-900$\,nm). Their spectral resolution at full slit is either 1000 or 5000 depending on the use of the low- or medium-resolution grating mode and on the camera positions.
During the initial \lamost\ observations, there was the possibility to enhance the resolution to 2000 or 10\,000 by limiting the fiber slit to 1/2 slit width for the low or medium grating mode. 
However, at the time of our observations, only one set of medium resolution gratings was available and the wavelength coverage of the medium resolution mode is restricted to $510-540$\,nm in the blue and $830-890$\,nm in the red. 
It was not used.
Moreover, at the start of the pilot survey on 2011 October 24, the fiber slit was permanently fixed to $\times$2/3 slit width for all the \lamost\ projects, corresponding to a resolution of about 1800 for the low resolution mode. The active optics technique is used to control the reflecting corrector \citep{Su1998SPIE.3352...76S}. The \lamost\ has a quasi-meridian transit configuration capable of tracking the motion of celestial objects during about 4 hours while they are passing the meridian. 
For more detailed information about the \lamost, see \citet{Cui2012RAA....12.1197C} and \citet{Zhao2012RAA....12..723Z}.

The \lamost\ is an excellent instrument to perform spectroscopic follow-up for targets of the \kepler\ mission as it is capable of collecting low-resolution spectra for thousands of objects down to magnitude 17.8 simultaneously. 
In 2010, we initiated the ``\lamost-\kepler\ project'' (\project) to observe as many objects in the \kepler\ FoV as possible from the start of the test phase of the \lamost\ onward. This allows a homogeneous determination of both the stellar parameters and the spectral classification of the observed objects. 
Moreover, with low-resolution spectra it is possible to estimate the radial velocity (\vrad) and, in case of rapid rotation, the projected rotational velocity (\vsini) of the observed objects.

This paper is the first of a series that deals with the detailed description and the analysis of the spectra obtained for the \project\ in the Kepler FoV during the first round of observations, up to the end of the 2014 observation season.
The structure of this paper is as follows. 
In Section\,\ref{sect:target} we explain how we compiled a prioritized target list that was used as input for the \lamost\ observations. In Sections\,\ref{sect:pointings} and \ref{sect:fibers}, we justify our choice of the central positions of the different \lamost\ fields that we requested to be observed and how the selection of the fibers on the fields was done, respectively.
A detailed description of the observations themselves is given in Section\,\ref{sect:observations}. 
Section\,\ref{sect:reductions} gives an overview of the most important steps of the reduction procedure that resulted in the database of \lamost\ spectra that is described in Section\,\ref{sect:database}. 
The full results will be published in the subsequent papers of this series (Ren et al., in preparation; Frasca et al., in preparation; Gray \& Corbally, in preparation).
This paper ends with a discussion of the importance of the acquired \lamost\ spectra for the community involved in the \kepler\ research in Section\,\ref{sect:discussion}.

%%%%%%%%%%%%%%%%%%%%%%%%%%%%%%%%%%%%%%%%%%%%%%%%%%%%%%%%%%%%%%%%%%%%%%%%%%%%%%
\section{Selection of the targets}
\label{sect:target}
%%%%%%%%%%%%%%%%%%%%%%%%%%%%%%%%%%%%%%%%%%%%%%%%%%%%%%%%%%%%%%%%%%%%%%%%%%%%%%

We constructed a prioritized list of targets for the \project\ that was used to prepare the \lamost\ observations. 
We made use of the knowledge of objects within the \kepler\ FoV prior to the start of our \project, i.e. their position (right ascension $\alpha_{2000}$, declination $\delta_{2000}$), their brightness (the KIC magnitude \Kp\ for most of the objects; see Sect.\,\ref{sect:bright}), the availability of stellar parameters in the KIC (the effective temperature \teff, the surface gravity \logg\ and the metallicity \mh) and their scientific importance within the community involved in the \kepler\ research (see Sect.\,\ref{sect:type}).

\subsection{Type of targets}
\label{sect:type}

\begin{figure*}
\begin{center}
\includegraphics[width=135mm]{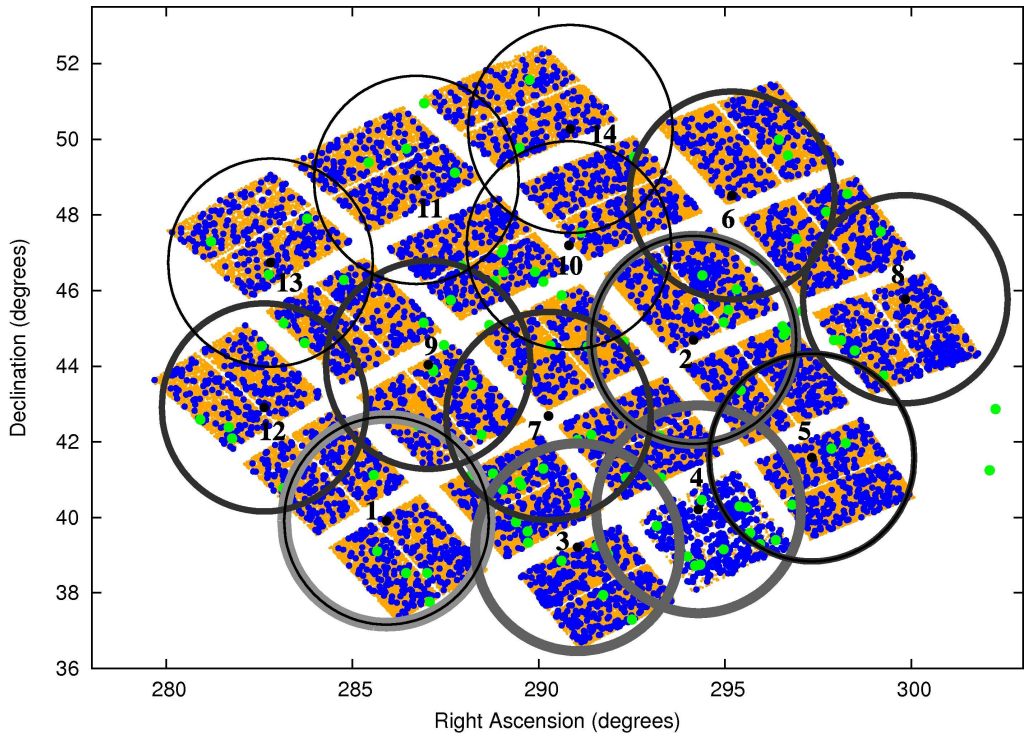}
\end{center}
\caption{\label{fig:LKfields} 
The targets of scientific interest in the FoV of the \kepler\ mission. The black dots refer to the centers of the 14 \pointings\ that cover the \kepler\ FoV (cf. Table\,\ref{tab:LKfields_ok}). 
The following color coding is used (from high to low scientific importance): green for standard targets, blue for \kasc\ targets and orange for planet targets. The scientific importance of the different types of targets within the \project\ is also reflected in the size of the symbols. 
The \pointings\ that have been observed in 2011, 2012, 2013, and 2014 are indicated by the circles drawn with a full line going from thick to thin and from gray to black, respectively. 
(The color representation is only available in the online version of the paper)
}
\end{figure*}

The first level of prioritization was based on the type of object. 
The targets with a specific scientific interest are represented in Fig.\,\ref{fig:LKfields}.
We made a distinction from high to low priority among: 

\begin{itemize}

\item ``standard targets'' ($\sim$120): 
MK secondary standard stars. 
These targets were originally introduced in the beginning of the \project\ because \lamost\ spectra of these objects are needed to calibrate the results of the other objects.
As the project evolved, other standard stars based on earlier \lamost\ observations were selected automatically.

\item  ``\kasc\ targets'' ($\sim$6\,500): 
Targets in the \kepler\ FoV selected by the \kepler\ Asteroseismic Science Consortium (\kasc).\footnote{Relevant information on \kasc\ can be found on the Kepler Asteroseismic Science Operations Center (KASOC) web page {\tt http://kasoc.phys.au.dk/}, maintained by Rasmus Handberg from the University of Birmingham in the United Kingdom.}.
As this project was initiated by the chairs of the ground-based follow-up working groups for \kasc\ targets, these \kasc\ targets were considered to be of the highest scientific interest.

\item ``planet targets'' ($\sim$150\,000):
 Targets in the \kepler\ FoV selected by the \kepler\ planet search group \citep{Batalha2010ApJ...713L.109B}. 

\item ``extra targets'' ($\sim$1\,000\,000): other targets in the \kepler\ FoV from the KIC \citep{Brown2011AJ....142..112B}. 
They have no specific scientific interest but were potential targets of the \kepler\ mission, except those that fall in the regions that are not covered by the \kepler\ CCDs (white parts on Fig.\,\ref{fig:LKfields}). 

\end{itemize}

From here on, we will refer to all these targets as \kepler\ targets. 
Note that there is overlap with the targets of the \lamost\ project of Wang et al. (private communication) focusing on planet host star candidates in the \kepler\ FoV. 
The targets of common interest received the same priority as the \kasc\ targets from 2012 onward.
It was already clear from the beginning of our project that not all the \kepler\ targets could be observed. Initially, the aim was to observe as many of the \kasc\ and planet targets as possible, as those targets are of a specific scientific interest. Now that the observations of the nominal \kepler\ mission have ended, we know for which objects there is \kepler\ photometry available. For the next round of observations, these 199\,718 objects will get the top priority.

To ensure an optimal use of all the available fibers, we provided for each of the requested \lamost\ fields (see Sect.\,\ref{sect:pointings}) a list of additional objects with a $V < 20$ based on the USNO-B catalog \citep{Monet2003AJ....125..984M}. 
These ``field targets'' are of no particular scientific interest for our project, but since they are observable and used to fill all the available fibers, we will get relevant information on them by analyzing their spectra homogeneously with the \kepler\ targets.
Note that the field targets have not yet been cross-identified with the KIC when compiling the target list.
This was only done afterwards for all the objects that were observed with the \lamost\ (see Sect.\,\ref{sect:identification}).
Of the 38\,416 objects that were originally observed as a field target, only 79 could not be reclassified as a \kepler\ target after cross-identification (cf. ``Unique'' row in Table\,\ref{tab:LKfields_ok}).

\subsection{Brightness} 
\label{sect:bright}

Ground-based spectroscopic data are also being collected with 2-m class telescopes elsewhere (e.g. \citealt{Uytterhoeven2010AN....331..993U,Molenda2010AN....331..981M,Catanzaro2010A&A...517A...3C,Bruntt2011A&A...528A.121B,Catanzaro2011MNRAS.411.1167C,Lehmann2011A&A...526A.124L,Molenda2011MNRAS.412.1210M,Bruntt2012MNRAS.423..122B,Thygesen2012A&A...543A.160T,Tkachenko2012MNRAS.422.2960T,Tkachenko2013MNRAS.431.3685T,Molenda2013MNRAS.434.1422M,Karoff2013MNRAS.433.3227K,Niemczura2015MNRAS.450.2764N}). 
As these observations are restricted to bright targets (generally $V < 12$), the highest priority for observations with the \lamost\ is given to the fainter objects.
However, several bright stars studied in the aforementioned works were also targeted for the comparison purposes.

The exposure times for \lamost\ observations can only be optimized if the brightness range of the observed targets is not larger than 5 magnitudes. 
For the initial observations (2011 May 31, 2011 June 8, and 2012 June 4), we therefore made a distinction between three different brightness intervals from high to low priority: middle targets ($10.5 < K{\rm p} \leq 15.5$), bright targets ($K{\rm p} \leq 10.5$), and faint targets ($K{\rm p} > 15.5$). 
For the group of middle targets, those without KIC stellar parameters are considered as the most important ones within each type of target (standard, \kasc, planet and extra). 
They were prioritized according to brightness from bright to faint. 
Those for which KIC stellar parameters are already available were prioritized from high to low \teff\ because the determination of the KIC \teff\ for hot stars suffered from the lack of an ultraviolet filter in the Sloan photometry. 
The objects in the other brightness groups were prioritized according to brightness regardless of the availability of KIC parameters. 
The bright targets were prioritized from faint to bright to minimize saturation effects (too much flux) while the faint targets were prioritized from bright to faint to minimize the occurrence of underexposed objects (not enough flux). 
For the observations obtained after 2012 June 4, only a distinction between brighter ($9.0 < K{\rm p} \leq 14.0$) and fainter ($K{\rm p} > 14.0$) targets was made. The brighter and fainter targets were prioritized in the same way as the middle and faint targets from the initial observations, respectively.

%%%%%%%%%%%%%%%%%%%%%%%%%%%%%%%%%%%%%%%%%%%%%%%%%%%%%%%%%%%%%%%%%%%%%%%%%%%%%%
\section{Selection of the LAMOST fields}
\label{sect:pointings}
%%%%%%%%%%%%%%%%%%%%%%%%%%%%%%%%%%%%%%%%%%%%%%%%%%%%%%%%%%%%%%%%%%%%%%%%%%%%%%

\begin{table*}
\begin{center}
\caption{\label{tab:LKfields_ok} 
Statistical overview of the observations that have been obtained up to the end of the 2014 observations season for the \kepler\ FoV for the \project.
}
  \begin{tabular}{lllcccrrrrrrr}
    \hline
    \pointing& RA(2000)    & DE(2000)     & Cluster & Date   &\# &  Spectra &     S &      K &       P &       E &       F &      KO \\
\hline                                                                                                                                      
    LK01    & 19:03:39.258 & +39:54:39.24 &         & 110530 & 1 &   1\,435 &     1 &    131 &     444 &     859 &         &     607 \\
            &              &              &         & 110608 & 2 &   1\,108 &     4 &    137 &     336 &     629 &       2 &     525 \\
            &              &              &         & 140602 & 2 &   4\,944 &     2 &    237 &  2\,201 &  2\,500 &       4 &  2\,586 \\
    LK02    & 19:36:37.977 & +44:41:41.77 & NGC6811 & 120604 & 1 &      582 &    18 &    131 &     157 &     273 &       3 &     355 \\
            &              &              &         & 140913 & 2 &   6\,858 &     5 &    483 &  3\,504 &  2\,861 &       5 &  4\,171 \\
    LK03    & 19:24:09.919 & +39:12:42.00 & NGC6791 & 120615 & 3 &   9\,655 &    18 &    591 &  4\,825 &  4\,215 &       6 &  5\,833 \\
    LK04    & 19:37:08.862 & +40:12:49.63 & NGC6819 & 120617 & 3 &   9\,666 &    18 &    574 &  3\,785 &  5\,281 &       8 &  5\,282 \\
    LK05    & 19:49:18.139 & +41:34:56.85 &         & 131005 & 2 &   6\,374 &     4 &    474 &  2\,704 &  3\,191 &       1 &  3\,487 \\
            &              &              &         & 140522 & 1 &   2\,422 &     4 &    278 &  1\,048 &  1\,092 &         &  1\,431 \\
    LK06    & 19:40:45.383 & +48:30:45.10 &         & 130522 & 1 &   2\,488 &     4 &    366 &  1\,383 &     735 &         &  1\,770 \\
            &              &              &         & 130914 & 1 &   3\,037 &     3 &    237 &  1\,778 &  1\,015 &       4 &  2\,046 \\
    LK07    & 19:21:02.816 & +42:41:13.07 &         & 130519 & 1 &   3\,025 &    12 &    416 &  1\,820 &     777 &         &  2\,303 \\
            &              &              &         & 130523 & 1 &   2\,304 &     4 &    152 &  1\,590 &     558 &         &  1\,793 \\
            &              &              &         & 130926 & 1 &   3\,165 &     8 &    275 &  1\,995 &     886 &       1 &  2\,357 \\
    LK08    & 19:59:20.425 & +45:46:21.15 & NGC6866 & 130925 & 2 &   5\,928 &     2 &    387 &  1\,957 &  3\,576 &       6 &  2\,386 \\
            &              &              &         & 131002 & 1 &   2\,795 &       &     20 &      19 &  2\,751 &       5 &      56 \\
            &              &              &         & 131017 & 1 &   2\,769 &       &     41 &     906 &  1\,820 &       2 &     990 \\
            &              &              &         & 131025 & 1 &   2\,970 &       &     50 &  1\,024 &  1\,896 &         &  1\,119 \\
    LK09    & 19:08:08.340 & +44:02:10.88 &         & 131004 & 1 &   3\,141 &     4 &    266 &  1\,748 &  1\,122 &       1 &  2\,106 \\
    LK10    & 19:23:14.829 & +47:11:44.80 &         & 140520 & 2 &   3\,781 &     4 &    222 &  2\,401 &  1\,154 &         &  2\,683 \\
    LK11    & 19:06:51.499 & +48:55:31.77 &         & 140918 & 1 &   3\,070 &     4 &    340 &  1\,572 &  1\,149 &       5 &  1\,940 \\
    LK12    & 18:50:31.041 & +42:54:43.72 &         & 131007 & 1 &   2\,922 &     5 &    229 &  1\,329 &  1\,354 &       5 &  1\,590 \\
    LK13    & 18:51:11.993 & +46:44:17.52 &         & 140502 & 1 &   2\,501 &     2 &    199 &  1\,102 &  1\,196 &       2 &  1\,345 \\
            &              &              &         & 140529 & 2 &   4\,693 &     1 &    232 &  2\,091 &  2\,358 &      11 &  2\,434 \\
    LK14    & 19:23:23.787 & +50:16:16.64 &         & 140917 & 1 &   3\,167 &       &    187 &  1\,455 &  1\,521 &       4 &  1\,685 \\
            &              &              &         & 140927 & 1 &   3\,160 &     1 &    107 &  1\,424 &  1\,625 &       3 &  1\,585 \\
            &              &              &         & 140929 & 1 &   3\,126 &     1 &     87 &  1\,291 &  1\,740 &       7 &  1\,441 \\
 \hline                                                                                                                                             
  Total     &            &               &         &         &   & 101\,086 &   129 & 6\,849 & 45\,889 & 48\,134 &      85 & 55\,906 \\
  Unique    &            &               &         &         &   &  80\,447 &    93 & 4\,712 & 34\,904 & 40\,659 &      79 & 42\,209 \\
  1$\times$ &            &               &         &         &   &  63\,333 &    65 & 3\,142 & 25\,919 & 34\,133 &      74 & 31\,147 \\
  2$\times$ &            &               &         &         &   &  14\,186 &    20 & 1\,129 &  7\,286 &  5\,747 &       4 &  8\,859 \\  
  3$\times$ &            &               &         &         &   &   2\,483 &     8 &    355 &  1\,457 &     662 &       1 &  1\,870 \\  
  4$\times$ &            &               &         &         &   &      332 &       &     54 &     199 &      79 &         &     258 \\  
 +5$\times$ &            &               &         &         &   &      113 &       &     32 &      43 &      38 &         &      75 \\  
 \hline 
\end{tabular}
\end{center}
The top lines give the specifications of the \pointings\ that have been observed. 
For each \pointing, we give the right ascension (RA(2000)) and declination (DE(2000)) of the central bright star, the name of the open cluster that it contains (Cluster, if applicable), the date of observation (YYMMDD; Date), the number of plates that were used to observe the \pointing\ (\#), the total number of spectra that have been reduced successfully (Spectra) and the number of spectra for \kepler\ targets - subdivided into standard (S), \kasc\ (K), planet (P) extra (E) -, for field targets (F) and for objects that were observed by the \kepler\ mission (KO).
The bottom lines give the summary of the observations of all \pointings\ together.
For each category of targets, we give the total number of spectra that have been reduced successfully (Total), the number of different objects that it contains (Unique) and the number of targets that have been observed one (1$\times$), two (2$\times$), three (3$\times$), four (4$\times$) and at least five (+5$\times$) times.
\end{table*}

We made use of a simplified version of the bright central star tiling method to cover the \kepler\ FoV \citep{Cui2012RAA....12.1197C}. 
This method requires a star brighter than $V = 8$  in the center of the \lamost\ FoV (for the active optics wave front sensor; central hole on Fig.\,\ref{fig:fov}) and four fainter stars with magnitude $V < 17$ (for the guiding of the CCD cameras; off-center holes on Fig.\,\ref{fig:fov}). 
We therefore started by selecting 14 circular regions with a diameter of 5\degree, which is the minimum to (almost) fully cover the \kepler\ FoV. 
We refer to them as \pointings. 

In a second step, we searched for the optimized central coordinates ($\alpha_m$, $\delta_m$) for the \pointings\ that maximize the total number of observable \kasc\ targets. 
This was done by adding/subtracting 0, 1 or 2 steps of 5\,arcsec to/from the central coordinates of the original \pointings. 
Finally, we searched for the bright stars with a $V < 8$ that are the closest to the resulting $\alpha_m$ and $\delta_m$ values. 
These bright stars are indicated by the numbers on Fig.\,\ref{fig:LKfields}. 
The specifications of the corresponding final \pointings\ are given in Table\,\ref{tab:LKfields_ok}.
The \pointings\ are given in the suggested observation order that was chosen to maximize the number of observable \kasc\ targets. 
Note that four of the \pointings\ contain an open cluster (NGC\,6791, NGC\,6811, NGC\,6819, NGC\,6866). 
Since members of open clusters have more constraints on input parameters that can be used for asteroseismic modeling (such as a common age and metallicity), the priority of the \pointings\ containing an open cluster has been increased according to the asteroseismic potential of the open cluster. 
Unfortunately, because of the high density of objects in these regions, it is impossible to observe all targets of scientific interest in the \pointings\ containing an open cluster.

%%%%%%%%%%%%%%%%%%%%%%%%%%%%%%%%%%%%%%%%%%%%%%%%%%%%%%%%%%%%%%%%%%%%%%%%%%%%%%
\section{Selection of the fibers}
\label{sect:fibers}
%%%%%%%%%%%%%%%%%%%%%%%%%%%%%%%%%%%%%%%%%%%%%%%%%%%%%%%%%%%%%%%%%%%%%%%%%%%%%%

The \lamost\ is equipped with 4\,000 optical fibers. All fibers have a diameter of 320\,$\mu$m corresponding to 3.3\,arcsec on the sky \citep{Cui2012RAA....12.1197C}. A new system for the positioning of the fibers, called the ``distributive parallel-controllable fiber positioning system'', has been developed \citep{Xing1998SPIE.3352..839X}. The focal plane of the telescope with a diameter of 1.75\,m (5\degree) is covered by 4\,000 fiber positioning units whose configuration is shown in Fig.\,\ref{fig:fov}. Each unit covers a circular area with a diameter of 33\,mm (340\,arcsec) on the focal surface and contains one fiber that can reach all places within this area \citep{Cui2012RAA....12.1197C}. There is some overlap between the areas to ensure that there is no blind space on the focal surface. The 4\,000 fibers can all move separately and the accurate positioning of the fibers only takes about 10 min. The total preparation time for an exposure, including the focus procedure, is about half an hour.

\begin{table*}
  \begin{center}
  \caption{\label{tab:LKweather} 
Overview of the specifications of the observed plates for the \project\ from 2011 to 2014.  
}
  \begin{tabular}{ccrlccll}
    \hline
    Plate&\pointing&\multicolumn{1}{c}{$\Delta$T$_{\rm tot}$}&\multicolumn{1}{c}{$\Delta$T$_{\rm sub}$}&Slit&Seeing&Transparency& Observer\\
         &         &\multicolumn{1}{c}{(s)}                &\multicolumn{1}{c}{(s)}               &    &(arcsec)&           &         \\ \hline
    110530\_1  & LK01 & 2\,400 &  2$\times$1\,200            &  $\times$1   & $\sim$2.75 & mostly clear    & Wei Zhang     \\ 
    110608\_1  & LK01 &    900 &  1$\times$600+1$\times$300  &  $\times$1/2 & $\sim$3.60 & clear           & Jianrong Shi  \\ 
    110608\_2  & LK01 & 1\,200 &  2$\times$600               &  $\times$1/2 & $\sim$3.60 & clear           & Jianrong Shi  \\ 
    120604\_1  & LK02 &    600 &  2$\times$300               &  $\times$2/3 & $\sim$3.10 & poor            & Huoming Shi   \\ 
    120615\_1  & LK03 & 1\,600 &  2$\times$600+1$\times$400  &  $\times$2/3 & $\sim$3.75 & mostly cloudy   & Jianrong Shi  \\ 
    120615\_2  & LK03 & 1\,300 &  2$\times$400+1$\times$500  &  $\times$2/3 & $\sim$3.25 & mostly cloudy   & Jianrong Shi  \\ 
    120615\_3  & LK03 & 2\,400 &  2$\times$1\,200            &  $\times$2/3 & $\sim$3.75 & mostly cloudy   & Jianrong Shi  \\ 
    120617\_1  & LK04 &    900 &  1$\times$400+1$\times$500  &  $\times$2/3 & $\sim$4.45 & cloudy          & Jianrong Shi  \\ 
    120617\_2  & LK04 &    900 &  1$\times$400+1$\times$500  &  $\times$2/3 & $\sim$3.90 & cloudy          & Jianrong Shi  \\ 
    120617\_3  & LK04 & 1\,200 &  1$\times$1\,200            &  $\times$2/3 & $\sim$4.00 & cloudy          & Jianrong Shi  \\ 
    130519\_1  & LK07 & 3\,000 &  5$\times$600               &  $\times$2/3 & $\sim$3.50 & clear           & Jianling Wang \\ %
    130522\_1  & LK06 & 2\,100 &  3$\times$700               &  $\times$2/3 & $\sim$3.25 & cloudy          & Jianling Wang \\ %
    130523\_1  & LK07 & 3\,600 &  4$\times$900               &  $\times$2/3 & $\sim$3.80 & cloudy          & Jianling Wang \\ %& (,) : & Kepler_2
    130914\_1  & LK06 & 1\,800 &  3$\times$600               &  $\times$2/3 & $\sim$3.10 & clear           & Jianling Wang \\ %& (295.189,48.5125) 20:40 & KP194045N483045V02
    130925\_1  & LK08 & 1\,800 &  3$\times$600               &  $\times$2/3 & $\sim$3.90 & clear           & Ming Yang     \\ %& (299.835,45.7725) 19:48 & KP195920N454621V01
    130925\_2  & LK08 & 1\,800 &  3$\times$600               &  $\times$2/3 & $\sim$4.10 & clear           & Ming Yang     \\ %& (299.835,45.7725) 20:43 & KP195920N454621V02
    130926\_1  & LK07 & 1\,800 &  3$\times$600               &  $\times$2/3 & $\sim$4.40 & clear           & Ming Yang     \\ %& (290.262,42.6870) 19:57 & KP192102N424113V02
    131002\_1  & LK08 & 4\,930 &2$\times$1\,800+1$\times$1\,330&$\times$2/3 & $\sim$3.90 & clear           & Wei Zhang     \\ %& (299.835,45.7725) 19:45 & KP195920N454621M01
    131004\_1  & LK09 & 1\,200 &  2$\times$600               &  $\times$2/3 & $\sim$3.60 & cloudy          & Jianrong Shi  \\ %& (287.035,44.0364) 19:45 & KP190808N440210V01
    131005\_1  & LK05 & 1\,800 &  3$\times$600               &  $\times$2/3 & $\sim$3.30 & cloudy          & Jianrong Shi  \\ %& (297.326,41.5825) 19:11 & KP194918N413456V01
    131005\_2  & LK05 & 1\,800 &  3$\times$600               &  $\times$2/3 & $\sim$3.50 & cloudy          & Jianrong Shi  \\ %& (297.326,41.5825) 20:07 & KP194918N413456V02
    131007\_1  & LK12 & 1\,800 &  3$\times$600               &  $\times$2/3 & $\sim$3.10 & clear           & Jianrong Shi  \\ %& (282.629,42.9121) 18:48 & KP185031N425443V01
    131017\_1  & LK08 & 1\,800 &  3$\times$600               &  $\times$2/3 & $\sim$3.30 & cloudy          & Jianling Wang \\ %& (299.835,45.7725) 19:17 & KP195920N454621V3
    131025\_1  & LK08 & 1\,800 &  3$\times$600               &  $\times$2/3 & $\sim$3.00 & clear           & Ming Yang     \\ %& (299.835,45.7725) 18:50 & KP195920N454621V3
    140502\_1  & LK13 & 2\,400 &  4$\times$600               &  $\times$2/3 & $\sim$3.50 & mostly clear    & Jianling Wang \\ %& (282.8  ,46.738 ) : & KP185111N464417V01
    140520\_1  & LK10 & 1\,500 &  1$\times$1\,500            &  $\times$2/3 & $\sim$3.20 & mostly clear    & Jianrong Shi  \\ %& (,) : & KP192314N471144B01
    140520\_2  & LK10 & 2\,400 &  4$\times$600               &  $\times$2/3 & $\sim$3.10 & mostly clear    & Jianrong Shi  \\ %& (,) : & KP192314N471144V01
    140522\_1  & LK05 & 1\,200 &  2$\times$600               &  $\times$2/3 & $\sim$3.40 & clear           & Jianrong Shi  \\ %& (297.326,41.58  ) 27:37 & KP194918N413456V01
    140529\_1  & LK13 & 1\,800 &  3$\times$600               &  $\times$2/3 & $\sim$3.70 & clear           & Wei Zhang     \\ %& (282.8  ,46.738 ) 25:24 & KP185111N464417V02
    140529\_2  & LK13 & 1\,800 &  3$\times$600               &  $\times$2/3 & $\sim$3.40 & clear           & Wei Zhang     \\ %& (282.8  ,46.738 ) 26:22 & KP185111N464417V03
    140602\_1  & LK01 & 1\,800 &  3$\times$600               &  $\times$2/3 & $\sim$3.10 & clear           & Ming Yang     \\ %& (285.914,39.9109) 25:44 & KP190339N395439V01
    140602\_2  & LK01 & 1\,800 &  3$\times$600               &  $\times$2/3 & $\sim$3.10 & clear           & Ming Yang     \\ %& (285.914,39.9109) 25:44 & KP190339N395439V02
    140913\_1  & LK02 & 1\,800 &  3$\times$600               &  $\times$2/3 & $\sim$3.00 & mostly clear    & Ming Yang \\ %& (       ,       ) 20:22 & KP193637N444141V01
    140913\_2  & LK02 & 1\,800 &  3$\times$600               &  $\times$2/3 & $\sim$3.20 & mostly clear    & Ming Yang \\ %& (       ,       ) 20:22 & KP193637N444141V02
    140917\_1  & LK14 & 1\,800 &  3$\times$600               &  $\times$2/3 & $\sim$3.20 & clear           & Ming Yang \\ %& (       ,       ) 20:32 & KP192323N501616V01
    140918\_1  & LK11 & 1\,800 &  3$\times$600               &  $\times$2/3 & $\sim$3.20 & clear           & Ming Yang \\ %& (       ,       ) 20:05 & KP190651N485531V01
    140927\_1  & LK14 & 1\,800 &  3$\times$600               &  $\times$2/3 & $\sim$3.80 & partly cloudy   & Jianrong Shi \\ %& (       ,       ) 19:34 & KP190651N485531V02
    140929\_1  & LK14 & 1\,800 &  3$\times$600               &  $\times$2/3 & $\sim$4.30 & clear           & Jianrong Shi \\ %& (       ,       ) 20:19 & KP192323N501616V03
    \hline 
  \end{tabular}
  \end{center}
For each plate, we give the date with the sequence number of the plate (YYMMDD\_N; Plate), the reference of the \pointing\ that was observed (\pointing, cf. Table\,\ref{tab:LKfields_ok}), the total integration time ($\Delta$T$_{\rm tot}$) and the integration times of the sub-exposures ($\Delta$T$_{\rm sub}$) in seconds, the slit width (Slit), the estimated value of the seeing (Seeing), an indication of the transparency of the sky during the observations (Transparency) and the name of the observer (Observer). 
\end{table*}

The code ``Survey Strategy System'' was used to prepare the observation plans of the \project\ to optimize the effective use of the fibers \citep{Cui2012RAA....12.1197C}. The prioritised list of targets (see Section\,\ref{sect:target}) and the equatorial coordinates of the selected bright central stars of the \pointings\ (see Section\,\ref{sect:pointings}) were used as input for the distribution of the fibers. During the allocation of the fibers, two basic rules apply: the number of observed objects should be maximised and objects that can be observed by more than one unit should be allocated to the unit that is the nearest to them to reduce the probability of mechanical collisions during the observations. For each spectrograph, at least five fibers should be allocated to standard targets whose flux is known to allow a flux calibration of the \lamost\ spectra, and twenty fibers should be pointing towards an object-free region to allow an accurate correction for the contribution of the sky. The displacement of a fiber from an object is determined by measuring the flux changes of objects passing through the fibers when the telescope points to different positions around the center of a sky field. The displacements of the 4\,000 fibers are used to build a coordinate relation between the focal surface and the celestial coordinates. For more details about these procedures, we refer to \citet{Cui2012RAA....12.1197C}.

%%%%%%%%%%%%%%%%%%%%%%%%%%%%%%%%%%%%%%%%%%%%%%%%%%%%%%%%%%%%%%%%%%%%%%%%%%%%%%
\section{Observations}
\label{sect:observations}
%%%%%%%%%%%%%%%%%%%%%%%%%%%%%%%%%%%%%%%%%%%%%%%%%%%%%%%%%%%%%%%%%%%%%%%%%%%%%%

As only one \pointing\ was observed per night, covered by at maximum three different plates (= unique configurations of the fibers), a strict minimum of 14 observation nights was required to observe all the requested \pointing s of the \project.
This is due to both the limitations of the \lamost\ and the visibility of the \kepler\ FoV. 
The observation of a \pointing\ can last at maximum 4\,h while the celestial objects are passing the meridian.
However, as the observable period of the \kepler\ FoV coincides for a large part with the Monsoon (the Xinglong observatory is closed from mid-June to mid-September), the observations can only be done from late-May to mid-June (with a maximal observation window of 2h before meridian passage) and from mid-September to late-October (with a maximal observation window of 2h after meridian passage).
About half an hour is needed to focus the system and CCD readout takes 8 min.
In general, 30\,min (3$\times$600\,s) are required to observe a ``V-plate'' (containing objects with a magnitude range of $9 < r \leq 14$) and 75\,min (3$\times$1\,500\,s) for a ``B-plate'' (containing objects with a magnitude range of $14 < r \leq 16.3$) (cf. Table\,\ref{tab:LKweather} from 2013 onward). 
For these reasons, four observation seasons of the \kepler\ FoV were needed to finish the first round of observations for the \project.

A total of 38 plates were observed on 27 different observation nights.
The earliest observations (in 2011 May and June), were done during the test phase of the \lamost. 
The 2012 observations were obtained during the pilot survey of \lamost\ (from 2011 October 24 until 2012 June 24).
The more recent observations date from the general survey phase of the \lamost. 
The \pointings\ that have been observed in the period 2011--2014 are indicated on Fig.\,\ref{fig:LKfields} by the circles drawn with a full line going from thick to thin and from gray to black, respectively. 
In Table\,\ref{tab:LKweather}, we give an overview of the specifications of the observed plates for the \project\ so far. 
In 2011, the resolution of the observations was either $R$\,= 1\,000 (slit position ``$\times$1'') or 2\,000 (slit position ``$\times$1/2'') while the resolution from 2012 onward was fixed to $R \sim 1\,800$ (slit position ``$\times$2/3''). 
The observation of a \lamost\ plate can last at maximum 4\,h as celestial objects can be observed from about 2\,h before to 2\,h after the passage at meridian.
The observation times of the sub-exposures of the plates were adjusted according to the brightness of the targets and to the weather conditions.
From 2013 on-wards, the observed plates are called a ``V-plate'', ``B-plate'', ``M-plate'' and ``F-plate'' for targets with $9 < r \leq 14$, $14 < r \leq 16.3$, $16.3 < r \leq 17.8$, and $17.8 < r \leq 18.5$, respectively.
Under normal circumstances, their observation times are 3 x 10 min, 3 x 25 min, 3 x 30 min, and 3 x 30 min, respectively ($\Delta$T$_{\rm sub}$ in Table\,\ref{tab:LKweather}). 

Before the Monsoon break of 2011, there were system errors between the astronomical coordinate system and the physical coordinates on the focal plane leading to many low-quality spectra. 
Thanks to a calibration of the system coordinates by using the stellar objects on the sky, the whole system error could be reduced to less than 0.5\,arcsec. Currently, the positioning errors can be compensated in real time ensuring a positioning accuracy of less than 40 microns, i.e. about 0.4\,arcsec \citep{Cui2012RAA....12.1197C}.

Before the pilot survey, there were occasional failures of the dewar system because of an exhaustion of the liquid nitrogen. Images taken under these circumstances are dominated by thermal noise due to the rising temperature of the CCD chip, making data obtained under these conditions useless. Moreover, there are about 110 known bad fibers either because they were broken or their fiber positioners do not work properly. Those fibers were excluded from the reduction pipeline.

%%%%%%%%%%%%%%%%%%%%%%%%%%%%%%%%%%%%%%%%%%%%%%%%%%%%%%%%%%%%%%%%%%%%%%%%%%%%%%
\section{Reductions}
\label{sect:reductions}
%%%%%%%%%%%%%%%%%%%%%%%%%%%%%%%%%%%%%%%%%%%%%%%%%%%%%%%%%%%%%%%%%%%%%%%%%%%%%%

\begin{table*}
  \begin{center}
  \caption{\label{tab:LKspectra} 
Overview of the \lamost\ 1D fits files that are available from the observations for the \project\ from 2011 to 2014. 
}
  \begin{tabular}{cclccccccccccccccccc}
    \hline
    Plate     & LJDN    & planID             & Version& \multicolumn{16}{c}{Spectrographs}                                            \\ %& Version
              &($\star$)&                    &        &                                                                               \\ %&
    \hline
    110530\_1  & 55712 & IF10M               & v2.6.5 & 01 & 02 & 03 & 04 & 05 & 06 &    &    &    &    &    &    & 13 &    &    &    \\ %& v2.6.5
    110608\_1  & 55721 & IF10B               & v2.6.5 & 01 &    & 03 & 04 & 05 & 06 & 07 &    &    &    &    &    &    &    & 15 &    \\ %& v2.4.4/v2.6.5
    110608\_2  & 55721 & IF10M               & v2.4.4 &    &    &    &    &    & 06 &    &    &    & 10 &    &    & 13 &    & 15 &    \\ %& v2.4.4
    120604\_1  & 56083 & IF04\_B56083        & v2.7.5 & 01 &    & 03 &    &    &    & 07 & 08 & 09 &    & 11 & 12 & 13 & 14 & 15 & 16 \\ %& v2.4.3/v2.7.5
    120615\_1  & 56094 & kepler05B56094      & v2.7.5 & 01 & 02 & 03 & 04 & 05 & 06 & 07 &    & 09 & 10 & 11 & 12 & 13 & 14 & 15 & 16 \\ %& v2.4.3/v2.7.5
    120615\_2  & 56094 & kepler05B56094\_2   & v2.7.5 & 01 & 02 & 03 & 04 & 05 & 06 & 07 &    & 09 & 10 & 11 & 12 & 13 & 14 & 15 & 16 \\ %& v2.4.3/v2.7.5
    120615\_3  & 56094 & kepler05F56094      & v2.7.5 & 01 & 02 & 03 &    & 05 & 06 & 07 &    & 09 & 10 & 11 & 12 & 13 & 14 & 15 & 16 \\ %& v2.4.3/v2.6.4/v2.7.5
    120617\_1  & 56096 & kepler08B56096\_1   & v2.7.5 & 01 & 02 & 03 &    & 05 & 06 & 07 & 08 & 09 &    & 11 & 12 & 13 & 14 & 15 & 16 \\ %& v2.4.3/v2.7.5
    120617\_2  & 56096 & kepler08B56096\_2   & v2.7.5 & 01 & 02 & 03 &    & 05 & 06 & 07 & 08 & 09 & 10 & 11 & 12 & 13 & 14 & 15 & 16 \\ %& v2.4.3/v2.7.5
    120617\_3  & 56096 & kepler08F56096      & v2.7.5 & 01 & 02 & 03 &    & 05 & 06 & 07 & 08 & 09 & 10 &    & 12 & 13 & 14 & 15 & 16 \\ %& v2.4.3/v2.7.5
    130519\_1  & 56432 & KP192102N424113V01  & v2.7.5 & 01 & 02 & 03 & 04 & 05 & 06 & 07 & 08 & 09 & 10 & 11 & 12 & 13 & 14 & 15 & 16 \\ %& v2.6.4/v2.7.5
    130522\_1  & 56435 & KP194045N483045V01  & v2.7.5 & 01 &    & 03 & 04 & 05 & 06 & 07 & 08 & 09 &    & 11 & 12 & 13 & 14 & 15 & 16 \\ %& v2.6.4/v2.7.5
    130523\_1  & 56436 & kepler02\_1         & v2.7.5 & 01 & 02 & 03 &    & 05 & 06 & 07 & 08 & 09 & 10 & 11 & 12 & 13 & 14 & 15 & 16 \\ %& v2.7.5
    130914\_1  & 56550 & KP194045N483045V02  & v2.7.5 &    & 02 & 03 & 04 & 05 &    & 07 & 08 & 09 & 10 & 11 & 12 & 13 & 14 & 15 & 16 \\ %& v2.6.5/v2.7.5
    130925\_1  & 56561 & KP195920N454621V01  & v2.7.5 & 01 & 02 & 03 & 04 & 05 & 06 & 07 & 08 & 09 & 10 & 11 & 12 & 13 & 14 & 15 &    \\ %& v2.6.5/v2.7.5
    130925\_2  & 56561 & KP195920N454621V02  & v2.7.5 & 01 & 02 & 03 & 04 & 05 & 06 & 07 & 08 & 09 & 10 & 11 & 12 & 13 & 14 & 15 &    \\ %& v2.6.5/v2.7.5
    130926\_1  & 56562 & KP192102N424113V02  & v2.7.5 & 01 & 02 & 03 & 04 & 05 & 06 & 07 & 08 & 09 & 10 & 11 & 12 & 13 & 14 & 15 & 16 \\ %& v2.6.5/v2.7.5
    131002\_1  & 56568 & KP195920N454621M01  & v2.7.5 & 01 & 02 & 03 & 04 & 05 &    & 07 & 08 & 09 & 10 & 11 & 12 &    & 14 & 15 & 16 \\ %& v2.6.5/v2.7.5
    131004\_1  & 56570 & KP190808N440210V01  & v2.7.5 & 01 & 02 & 03 & 04 & 05 & 06 & 07 & 08 & 09 & 10 & 11 & 12 & 13 & 14 & 15 & 16 \\ %& v2.6.5/v2.7.5
    131005\_1  & 56571 & KP194918N413456V01  & v2.7.5 & 01 & 02 & 03 & 04 & 05 & 06 & 07 & 08 & 09 & 10 & 11 & 12 & 13 & 14 & 15 & 16 \\ %& v2.6.5/v2.7.5
    131005\_2  & 56571 & KP194918N413456V02  & v2.7.5 & 01 & 02 & 03 & 04 & 05 & 06 & 07 & 08 & 09 & 10 & 11 & 12 & 13 & 14 & 15 & 16 \\ %& v2.6.5/v2.7.5
    131007\_1  & 56573 & KP185031N425443V01  & v2.7.5 & 01 & 02 & 03 & 04 & 05 &    & 07 & 08 & 09 & 10 & 11 & 12 & 13 & 14 & 15 & 16 \\ %& v2.6.5/v2.7.5
    131017\_1  & 56583 & KP195920N454621V03  & v2.7.5 &    & 02 & 03 & 04 & 05 &    & 07 & 08 & 09 & 10 & 11 & 12 & 13 & 14 & 15 & 16 \\ %& v2.6.5/v2.7.5
    131025\_1  & 56591 & KP195920N454621V3   & v2.7.5 & 01 & 02 & 03 & 04 & 05 &    & 07 & 08 & 09 & 10 & 11 & 12 & 13 & 14 & 15 & 16 \\ %& v2.6.5/v2.7.5
    140502\_1  & 56780 & KP185111N464417V01  & v2.7.5 & 01 & 02 & 03 & 04 & 05 & 06 & 07 & 08 & 09 & 10 &    & 12 &    & 14 & 15 & 16 \\ %& v2.7.5
    140520\_1  & 56798 & KP192314N471144B01  & v2.7.5 & 01 &    & 03 & 04 & 05 & 06 &    &    &    & 10 & 11 & 12 &    &    & 15 &    \\ %& v2.7.5
    140520\_2  & 56798 & KP192314N471144V01  & v2.7.5 & 01 & 02 & 03 & 04 & 05 & 06 & 07 & 08 & 09 & 10 & 11 & 12 & 13 & 14 & 15 & 16 \\ %& v2.7.5
    140522\_1  & 56800 & KP194918N413456V    & v2.7.5 & 01 & 02 & 03 & 04 & 05 & 06 & 07 & 08 & 09 & 10 & 11 &    & 13 & 14 & 15 & 16 \\ %& v2.7.5
    140529\_1  & 56807 & KP185111N464417V02  & v2.7.5 & 01 & 02 & 03 & 04 & 05 & 06 & 07 & 08 & 09 & 10 &    & 12 & 13 & 14 & 15 &    \\ %& v2.7.5
    140529\_2  & 56807 & KP185111N464417V03  & v2.7.5 & 01 & 02 & 03 & 04 & 05 & 06 & 07 & 08 & 09 & 10 & 11 & 12 & 13 & 14 & 15 &    \\ %& v2.7.5
    140602\_1  & 56811 & KP190339N395439V01  & v2.7.5 & 01 & 02 & 03 & 04 & 05 & 06 & 07 &    & 09 &    &    & 12 & 13 & 14 & 15 & 16 \\ %& v2.7.5
    140602\_2  & 56811 & KP190339N395439V02  & v2.7.5 & 01 & 02 & 03 & 04 & 05 & 06 &    & 08 & 09 & 10 & 11 & 12 & 13 & 14 & 15 & 16 \\ %& v2.7.5
    140913\_1  & 56914 & KP193637N444141V01  & v2.7.5 & 01 & 02 & 03 & 04 & 05 & 06 & 07 & 08 & 09 & 10 & 11 & 12 & 13 & 14 & 15 & 16 \\ %& v2.7.5
    140913\_2  & 56914 & KP193637N444141V02  & v2.7.5 & 01 & 02 & 03 & 04 & 05 & 06 & 07 & 08 & 09 & 10 & 11 & 12 & 13 & 14 & 15 & 16 \\ %& v2.7.5
    140917\_1  & 56918 & KP192323N501616V    & v2.7.5 & 01 & 02 & 03 & 04 & 05 & 06 & 07 &    & 09 & 10 & 11 & 12 & 13 & 14 & 15 & 16 \\ %& v2.7.5
    140918\_1  & 56919 & KP190651N485531V01  & v2.7.5 & 01 & 02 & 03 & 04 & 05 & 06 & 07 &    & 09 & 10 & 11 & 12 & 13 & 14 & 15 & 16 \\ %& v2.7.5
    140927\_1  & 56928 & KP192323N501616V02  & v2.7.5 & 01 & 02 & 03 & 04 &    & 06 & 07 & 08 & 09 & 10 & 11 & 12 & 13 & 14 & 15 & 16 \\ %& v2.7.5
    140929\_1  & 56930 & KP192323N501616V03  & v2.7.5 & 01 & 02 & 03 & 04 & 05 & 06 & 07 & 08 & 09 & 10 & 11 & 12 & 13 & 14 &    & 16 \\ %& v2.7.5
    \hline 
  \end{tabular}
  \end{center}
For each plate, we give the date with the sequence number of the plate (YYMMDD\_N; Plate), the \lamost\ modified Julian day number (LJDN; see note ($\star$) below for the definition) and the plan identification number (planID) as used in the name structure of the \lamost\ 1D fits files, the latest available version of the reduction pipeline (first two numbers) and analysis pipeline (last number) for which the results are available (Version), and spectrograph identification numbers for which data is available in the remaining columns (Spectrographs).
\\
($\star$) The \lamost\ modified Julian day number (LJDN) is the integer part of the \lamost\ modified Julian date (LJD) that is defined as modified Julian date (= JD - 2\,400\,000.5) the after adding 20 hours (5/6). This puts the start of the LJD timescale to 1858 November 16 at noon in local Beijing time. Hence, all the \lamost\ data obtained during one observation night will have the same LJDN.
\end{table*}

The main products of the \lamost\ are wavelength and flux calibrated spectra that are provided to the astronomers. They are processed by a data reduction pipeline and an analysis pipeline. In the following subsections, we give short descriptions of the most important reduction steps that are relevant for our project. The analysis pipeline includes the classification and identification of the spectra and the measurement of the spectral parameters of the observed objects. 
For more details about pipeline procedures, we refer to \citet{Luo2012RAA....12.1243L} and \citet{Luo2015arXiv150501570L}.

\subsection{Pipeline}

The data reduction pipeline consists of 2D and 1D procedures. The \lamost\ 2D procedure is used to reduce the data from each of the 32 CCD chips (2 for each spectrograph) and of each exposure separately. 
Apart from the science frames themselves, the 2D pipeline also makes use of the series of bias, dark, arc lamp and sky flat-field frames that are taken at the beginning of each observation night. 
The reduction procedure consists of several steps. 
The bias and dark frames are combined into a master bias and a master dark frame each night and they are both subtracted from each raw image. 
From the general survey onward, the dark subtraction is not necessary any more because the problem with an inside CCD light source that affected the images has been eliminated.
For each CCD chip, 250 1D spectra are extracted from the raw data by using the row positions derived from the flat-field frames. 
For each region of the sky covered by one spectrograph, a super sky based on the sky measurements of at least 20 fibers is created to model and subtract the contribution of the background. This step is important because the telluric continuum and lines, if not properly subtracted, affect the following analysis, especially for the faintest targets. The centroids of the lines of the arc lamp spectra are measured and a Legendre polynomial is fitted to their wavelengths as a function of pixel. 
A vacuum wavelength scale is applied for the wavelength calibration and the resulting wavelength calibrations are accurate to 10\,\kms\ (or better). 
Also the heliocentric corrections are applied. 
The difference in efficiency of the different fibers is corrected for by dividing with a super flat that is constructed by combining the extracted, wavelength-calibrated and re-scaled sky flat-field frames.
The telluric absorption lines are removed in four wavelength regions (685-696\,nm; 715-735\,nm; 756-772\,nm; 810-824\,nm). 
The spectra are flux-calibrated by matching the instrumental fluxes in the \lamost\ spectra of the observed standard targets with tabulated values in energetic units. The spectra from the red and blue arms of each spectrograph are combined by stacking the points with corresponding wavelengths using a B-spline function with inverse variance weighting. Outliers due to cosmic rays are rejected and masked and the errors in flux estimated. The combined spectra are re-sampled to constant velocity pixels with a pixel scale of 69\,\kms, which corresponds to a wavelength difference of $\Delta \log(\lambda) = 0.0001$.

In the \lamost\ 1D procedure, the reduced spectra from the \lamost\ 2D procedure are used to estimate the spectral class and the \vrad\ of the observed targets. 
These characteristics are based on template matching since for low signal-to-noise (SNR) spectra, the fitting and measurement of spectral lines is not precise. 
The \vrad\ determination is done with a generalized version of the cross-correlation technique. 
The analysis code estimates the spectral type (SpT) of the observed targets by making $\chi^2$ fits in wavelength space of their \lamost\ spectra to templates that were constructed from a set of Sloan Digital Sky Survey spectra (SDSS). 
The results are checked by comparison of the lines of the observed spectrum with those from the best-fitted template. 
If they are very different, it is decided by a visual check if the classification is correct or not. 
The 1D pipeline only gives results for spectra with a SNR\,$>$\,5. 
After comparison of the results with the spectral types and radial velocities of the observed SDSS stars, it turns out that 96\,percent of the spectra with SNR\,$>$\,10 in the $r$-band are classified correctly and an accuracy of 13\,\kms\ can be reached for the \vrad.

\subsection{Combination of the sub-exposures}

For the \lamost, the observation of each plate consists, in general, of more than one sub-exposure (see Table\,\ref{tab:LKweather}).
This observation strategy is preferred as it makes it possible both to subtract the cosmic rays from the CCDs and to improve the SNR by combining the flux of the sequential sub-exposures.
The wavelength and flux calibration are done for each sub-exposure individually to overcome the unstable effects of the instrument (such as tracking and pointing) and the temporal variations of the weather.
In a first step, these sub-spectra are scaled to the same level before putting them together.
As the wavelength sample of the sub-spectra are not exactly the same, the combined spectrum is over-sampled.
In a second step, this combined spectrum is smoothed and re-sampled by a spline function.
Note that if the same field was observed on different nights, the spectra are not summed because the calibration is different for each night and the observed object might be a variable.

%%%%%%%%%%%%%%%%%%%%%%%%%%%%%%%%%%%%%%%%%%%%%%%%%%%%%%%%%%%%%%%%%%%%%%%%%%%%%%
\section{Database}
\label{sect:database}
%%%%%%%%%%%%%%%%%%%%%%%%%%%%%%%%%%%%%%%%%%%%%%%%%%%%%%%%%%%%%%%%%%%%%%%%%%%%%%

\begin{table*}
  \begin{center}
  \caption{\label{tab:database1D} 
Database of the \lamost\ spectra obtained for the \project: information extracted from the headers of the \lamost\ 1D fits files.
}
  \begin{tabular}{lccccccc}
\hline                      
 (1)                       & (2)         & (3)                 & (4)                    & (5)   & (6)    & (7)  & (8)      \\
 File Name                 & Target      & ID                  & Obs                    & Type  & KO     & Slit & Version  \\
                           & (9)         & (10)                & (11)                   & (12)  & (13)   & (14)  \\
                           & RAdeg       & DEdeg               & MAGtype                & MAG   & \snrr  & Notes \\
                           & (\degree)   & (\degree)           &                        & (mag) &        &       \\
\hline
spec-55712-IF10M\_sp01-001 & KIC02140382 & ZHT17\_Std\_0000110 & 2011-05-30T19:01:19.98 &   E   &   N    &   x1 &  v2.6.5  \\
                           & 285.914888  &   37.548585         & SSA\_BR2I              & 14.95 &  49.81 &      \\
spec-55712-IF10M\_sp01-002 & KIC01995622 & extra00011944       & 2011-05-30T19:01:19.98 &  P    &  Y     &   x1 &  v2.6.5  \\
                           & 286.184208  &   37.482444         & Kp                     & 13.52 & 112.71 &      \\
spec-55712-IF10M\_sp01-003 & KIC01994759 & ZHT17\_Std\_0000095 & 2011-05-30T19:01:19.98 &   E   &   N    &   x1 &  v2.6.5  \\
                           & 285.926985  &   37.494571         & SSA\_BR2I              & 16.73 &   9.25 &      \\
spec-55712-IF10M\_sp01-004 & KIC02141025 & extra00014786       & 2011-05-30T19:01:19.98 &   E   &   N    &   x1 &  v2.6.5  \\
                           & 286.134417  &   37.560833         & Kp                     & 14.83 &  41.10 &      \\
spec-55712-IF10M\_sp01-005 & KIC01995904 & extra00012008       & 2011-05-30T19:01:19.98 &   E   &   N    &   x1 &  v2.6.5  \\
                           & 286.261583  &   37.482583         & Kp                     & 14.94 &  47.37 &      \\
spec-55712-IF10M\_sp01-006 & KIC02283124 & extra00018771       & 2011-05-30T19:01:19.98 &  P    &  Y     &   x1 &  v2.6.5  \\
                           & 286.034917  &   37.664778         & Kp                     & 12.62 & 202.07 &      \\
spec-55712-IF10M\_sp01-008 & KIC02283334 & extra00018819       & 2011-05-30T19:01:19.98 &  P    &  Y     &   x1 &  v2.6.5  \\
                           & 286.104708  &   37.604194         & Kp                     & 15.37 &  33.25 &      \\
spec-55712-IF10M\_sp01-009 & KIC02282652 & extra00018639       & 2011-05-30T19:01:19.98 &  P    &  Y     &   x1 &  v2.6.5  \\
                           & 285.862583  &   37.682472         & Kp                     & 15.01 &  51.25 &      \\
spec-55712-IF10M\_sp01-010 & KIC02283496 & extra00018853       & 2011-05-30T19:01:19.98 &  P    &  Y     &   x1 &  v2.6.5  \\
                           & 286.155583  &   37.667000         & Kp                     & 15.29 &  42.23 &      \\
spec-55712-IF10M\_sp01-011 & KIC02422998 & kplr002422998       & 2011-05-30T19:01:19.98 & K     &  Y     &   x1 &  v2.6.5  \\
                           & 286.192417  &   37.740083         & Kp                     & 10.63 & 531.39 &      \\
\vdots                     & \vdots      & \vdots              & \vdots                 & \vdots& \vdots &\vdots& \vdots   \\
\hline 
  \end{tabular}
  \end{center}
Only the first 10 entries are shown.
%The full table is available only in electronic form only via anonymous ftp to cdsarc.u-strasbg.fr (130.79.128.5) at the CDS.
The full table is available in electronic form only.
For a description of the columns, we refer to the text (see Sect.\,\ref{sect:description}).
For clarity, columns (9)--(14) are given on a second line for each entry.
\end{table*}

\begin{table*}
  \begin{center}
  \caption{\label{tab:databaseKIC} 
Database of the \lamost\ spectra obtained for the \project: additional information about the cross-identified \kepler\ targets extracted from the KIC (\citealt{Brown2011AJ....142..112B}).
}
  \begin{tabular}{lccccccc}
\hline                                                                                                                                        
(1)                        & (2)         & (3)          & (4)         & (5)        & (6)        & (7)             & (8)             \\ 
File Name                  & Target      &  RA          & DE          & Ang        & Con        & Kp              & MagDiff         \\ 
                           &             & (hh:mm:ss)   & (dd:mm:ss)  & (arcmin)   &            & (mag)           & (mag)           \\ 
(9)                        & (10)        & (11)         & (12)        & (13)       & (14)       & (15)            & (16)            \\
\teff                      & \logg       & \mh          & E(B-V)      & pm         & pmRA       & pmDE            & Notes           \\ 
(K)                        & (dex)       & (dex)        & (mag)       & (arcsec/yr)& (arcsec/yr)& (arcsec/yr)     &                 \\
\hline                                                                                                           
spec-55712-IF10M\_sp01-001 & KIC02140382 & 19:03:39.559 & 37:32:54.92 & 0.003      &            & 14.734          &   0.216         \\
                      6116 &       4.614 &       -0.319 &     0.127   &      0.017 & 0.015      &      0.007      &                 \\ 
spec-55712-IF10M\_sp01-002 & KIC01995622 & 19:04:44.213 & 37:28:56.78 & 0.0006     & 0.034(\#4) & 13.517          &   0.003         \\
                      5675 &       4.223 &       -0.748 &     0.108   &      0.004 & 0.003      &     -0.003      &                 \\ 
spec-55712-IF10M\_sp01-003 & KIC01994759 & 19:03:42.471 & 37:29:40.58 & 0.002      &            & 16.212          &   0.518         \\
                           &             &              &             &      0.000 & 0.000      &      0.000      &                 \\ 
spec-55712-IF10M\_sp01-004 & KIC02141025 & 19:04:32.263 & 37:33:38.95 & 0.001      & 0.066(\#4) & 14.827          &   0.003         \\
                      5916 &       4.294 &       -0.096 &     0.151   &      0.010 &-0.003      &     -0.010      &                 \\ 
spec-55712-IF10M\_sp01-005 & KIC01995904 & 19:05:02.777 & 37:28:57.32 & 0.0007     & 0.116(\#4) & 14.944          &  -0.004         \\
                      6048 &       4.281 &       -0.166 &     0.159   &      0.000 & 0.000      &      0.000      &                 \\ 
spec-55712-IF10M\_sp01-006 & KIC02283124 & 19:04:08.376 & 37:39:53.24 & 0.001      & 0.023(\#4) & 12.622          &  -0.002         \\
                      6821 &       4.232 &        0.045 &     0.114   &      0.011 &-0.002      &     -0.011      &                 \\ 
spec-55712-IF10M\_sp01-008 & KIC02283334 & 19:04:25.126 & 37:36:15.12 & 0.0009     & 0.099(\#4) & 15.372          &  -0.002         \\
                      6067 &       4.557 &       -0.170 &     0.151   &      0.004 & 0.0002     &     -0.004      &                 \\ 
spec-55712-IF10M\_sp01-009 & KIC02282652 & 19:03:27.017 & 37:40:56.93 & 0.0007     & 0.068(\#4) & 15.005          &   0.005         \\
                      4910 &       4.336 &        0.018 &     0.109   &      0.005 &-0.005      &      0.0003     &                 \\
spec-55712-IF10M\_sp01-010 & KIC02283496 & 19:04:37.339 & 37:40:01.20 & 0.0001     & 0.118(\#4) & 15.286          &   0.004         \\
                      5397 &       4.438 &       -0.170 &     0.133   &      0.010 & 0.009      &     -0.005      &                 \\ 
spec-55712-IF10M\_sp01-011 & KIC02422998 & 19:04:46.178 & 37:44:24.32 & 0.0005     & 0.006(\#4) & 10.630          &   0.000         \\
                      5185 &       4.644 &       -0.728 &     0.019   &      0.0005&-0.0001     &     -0.0003     &                 \\
\vdots                     & \vdots      & \vdots       & \vdots      & \vdots     & \vdots     & \vdots          & \vdots          \\
\hline 
  \end{tabular}
  \end{center}
Only the first ten entries are shown. 
%The full table is available only in electronic form only via anonymous ftp to cdsarc.u-strasbg.fr (130.79.128.5) at the CDS.
The full table is available in electronic form only.
For a description of the columns, we refer to the text (see Sect.\,\ref{sect:description}).
For clarity, columns (9)--(16) are given on a second line for each entry.
\end{table*}

A total of 38 plates have been observed so far for the \project.
Unfortunately, due to the pointing problems and/or the malfunctioning of some of the spectrographs and/or fibers (Sect.\,\ref{sect:observations}), the raw spectra for a fraction of the observed objects could not be reduced by the reduction pipeline for some of the nights, especially during the test phase and the pilot survey in 2011 and 2012 (Sect.\,\ref{sect:reductions}). 
This is reflected both in the amount of missing spectrograph numbers in Table\,\ref{tab:LKspectra} and in the evolution of the success rate of the reduction pipeline of the \lamost\ for the \project, as shown as the full line in light gray in Fig.\,\ref{fig:snrfiber}.
In the \project, only the \lamost\ observations for which both the 2D and 1D reduction procedures could be applied successfully are analyzed. 
We did not include in the database those spectra for which the 1D reduction procedure classified the object as either ``Unused'' (fiber is not used), ``Unassigned'' (fiber is not assigned), ``Dead'' (fiber is not working) or ``Sky'' (fiber is used to measure the flux of the sky).
This results in 101\,086 entries in the database for a total of 80\,447 different targets.
In Table\,\ref{tab:LKfields_ok}, we give the distribution of these targets for each of the observed \pointings. 
In Table\,\ref{tab:LKspectra}, we give an overview of the name structure of the files containing the \lamost\ spectra and of the spectrograph identification numbers for which \lamost\ data are available for the different plates of the different nights.

We have constructed a database for these 101\,086 \lamost\ spectra available for the \project\ so far (Tables\,\ref{tab:database1D} and \ref{tab:databaseKIC})\footnote{The full tables are available in electronic form only}.
The \lamost\ 1D fits files of the \lamost\ spectra, including those for the \project, that are already released to the public can be downloaded from the \lamost's official website\footnote{The \lamost\ 1D fits files of the released spectra of the \project\ can be downloaded from the \lamost's official website: {\tt http://www.lamost.org/}}.
The \lamost\ 1D fits files of the unreleased spectra that were made available to us by the \lamost\ team can be obtained upon request\footnote{The \lamost\ 1D fits files of the unreleased spectra of the \project\ can be requested by e-mail to Peter De Cat: Peter.DeCat@oma.be} after becoming an external collaborator of the \project\footnote{A request to become an external collaborator of the \project\ should be send by e-mail to Jianning Fu: jnfu@bnu.edu.cn}.
In column\,8 of Table\,\ref{tab:database1D}, we give the most recent version of the pipelines for which the results are available. The first two numbers refer to the reduction pipeline and the last number to the analysis pipeline.
Note that v2.7.5 refers to the most recent versions that are available at this moment. The spectra emerging from an older version of the reduction pipeline can be used at own risk.
The structure of the names of the \lamost\ 1D fits files is as follows: spec-MMMMM-YYYY\_spXX-FFF.fit where MMMMM gives the \lamost\ modified Julian day number (see note of Table\,\ref{tab:LKspectra} for the definition), YYYY denotes the plan identification number and XX and FFF are the spectrograph and fiber identification number, respectively. 
The \lamost\ modified Julian day number (LJDN) and the plan identification number (planID) of the observed plates are given in Table\,\ref{tab:LKspectra}.

\subsection{Cross-identification}
\label{sect:identification}

For all the available \lamost\ 1D fits files, we cross-checked if the observed objects could be identified with a \kepler\ object from the KIC. 
The right ascension and declination of the position of the fibers as recorded in the headers in the \lamost\ 1D fits files (see columns\,9 and 10 in Table\,\ref{tab:database1D}) were used as input for the KIC search tool\footnote{The \kepler\ Input Catalog can be consulted with the KIC search tool: {\tt http://archive.stsci.edu/kepler/kic10/search.php}}.
These coordinates generally correspond to the equatorial coordinates of the objects as given in the target list. For a fraction of the objects brighter than magnitude 11, an offset is added to the pointing of the fibers to avoid saturation. The term ``offset'' refers here to the difference between the equatorial coordinates of the position of the fibers and the coordinates of the observed object as given in the target list. The objects that were flagged as having a large offset are marked with ``N1'' in column\,14 of Table\,\ref{tab:database1D}.

We initially searched for objects within the default search radius of 0.02\,arcmin of these coordinates.
If no \kepler\ object was found in this region, we enlarged the search region to a maximum radius of 0.16 arcmin (= 10 arcsec).   
Objects for which the cross-identification was only found with such a larger search radius are marked with ``N2'' or ``N3'' in column\,14 of Table\,\ref{tab:database1D} and column\,16 of Table\,\ref{tab:databaseKIC}.
The few observations for which another KIC object matches better with the equatorial coordinates of the fiber than the \kepler\ target that is mentioned in the headers of the \lamost\ 1D fits files (column 3 of Table\,\ref{tab:database1D}) are marked with ``N4'' to ``N15'' in column\,14 of Table\,\ref{tab:database1D} and column\,16 of Table\,\ref{tab:databaseKIC}.
All the marked cross-identifications should be treated with caution.
In case an object could be cross-identified with a KIC object, we adopted the KIC\,number as the final identification of the object and retrieved the KIC values of the stellar parameters (\teff, \logg\ and \mh), the reddening (\ebv), the contamination factors and proper motions and listed them in the database, if available (see Table\,\ref{tab:databaseKIC}).

\begin{figure}
 \begin{center}
  \resizebox{0.4\textwidth}{!}{\includegraphics{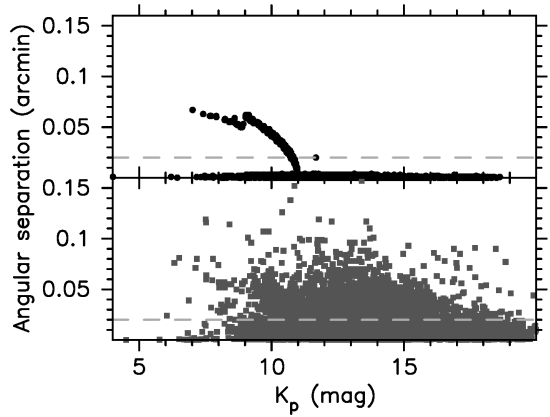}}
 \end{center}
\caption{\label{fig:angsep} 
Overview of the angular separation (in arcmin) between the equatorial coordinates of the fibers as given in the headers of the \lamost\ 1D fits files and the equatorial coordinates of the KIC objects they were cross-identified with (column\,5 of Table\,\ref{tab:databaseKIC}) as a function of their Kepler magnitude \Kp\ (column\,7 of Table\,\ref{tab:databaseKIC}). The observed objects that were listed as a KIC object in the target list are given on the top panel (black dots). The objects that were originally observed as a field target are given on the bottom panel (gray squares). The horizontal dashed lines in light gray indicate the standard search radius of 0.02\,arcmin around the equatorial coordinates of the fibers during the cross-identification process.
}
\end{figure}

An overview of the ``angular separation'' between the equatorial coordinates of these cross-identified objects and the equatorial coordinates of the position of their fiber (column\,5 of Table\,\ref{tab:databaseKIC}) as a function of their Kepler magnitude \Kp\ (column\,7 of Table\,\ref{tab:databaseKIC}) is given in Fig.\,\ref{fig:angsep}. The higher the angular separation, the more doubtful is the cross-identification. The horizontal dashed line indicates the standard search radius of 0.02\,arcmin around the equatorial coordinates of the position of the fibers. The top panel shows the observed objects that were listed as a KIC object in the target list. It clearly shows that the angular separations of these objects are minimal except for a fraction of the targets with magnitude \Kp\,$>$ 11. The amount of offset that has been added to the pointings of the fibers is related to the brightness of these objects. The bottom panel shows the objects that were originally observed as a field target but that were cross-identified with a KIC object. For these targets, the ``offset'' and ``angular separation'' are not expected to be identical as their coordinates given in the target list originate from the USNO-B Catalog \citep{Monet2003AJ....125..984M} and not from the KIC \citep{Brown2011AJ....142..112B}. Indeed, for these objects, large angular separations are found for objects of all magnitudes and no clear relation is found between the angular separation and the brightness of the observed objects. We recall that the cross-identifications of targets with a large angular separation should be treated with caution.

\subsection{Description}
\label{sect:description}

For all the \lamost\ spectra in the database, Table\,\ref{tab:database1D} lists information contained in or extracted from the headers of their 1D fits file. This table contains the following columns:

\begin{itemize}

\item column\,1: The file name of the \lamost\ 1D fits file (File Name).

\item column\,2: The final identification of the target after cross-identification with the KIC catalog (Target).

\item column\,3: The identification of the target as given in headers of the \lamost\ 1D fits file (ID).

\item column\,4: The date and time (Obs) of mid-exposure (yyyy-mm-ddThh:mm:ss.ss).

\item column\,5: The type of the target (Type). The standard targets are indicated with ``S'', \kasc\ targets with ``K'', planet targets with ``P'', extra targets with ``E'' and field targets with ``F''.

\item column\,6: The availability of data from the \kepler\ mission for this object (KO). ``Y'' means that the object has been observed by the \kepler\ mission and hence that \kepler\ data is available. ``N'' means no \kepler\ data is available.

\item column\,7: The position of the slit (Slit). 
``$\times$1'' means that the fiber slit uses the full slit width ($R$\,= 1000). In case of ``$\times$2/3'' or ``$\times$1/2'', the fiber slit is limited to 2/3 ($R$\,$\sim$ 1800) or half ($R$\,= 2000) of the slit width, respectively.

\item column\,8: The most recent version number of the pipelines for which the results are available (Version). The first two numbers refer to the reduction pipeline and the last number to the analysis pipeline.

\item column\,9: The right ascension of the fiber in degrees (RAdeg). In general, it corresponds to the right ascension of the object as given in the target list (epoch 2000.0) except for a fraction of the targets brighter than magnitude 11, for which an offset is added to the pointings of the fibers to avoid saturation.

\item column\,10: Idem as column\,9, but for the declination of the fiber in degrees (DEdeg).

\item column\,11: The type of magnitude as given in headers of the \lamost\ 1D fits file (MAGtype).

\item column\,12: The brightness of the target in magnitude as given in headers of the \lamost\ 1D fits file (MAG). Targets for which the brightness is not known have no entry in this column. 

\item column\,13: The signal-to-noise ratio in the r-band (\snrr). The entry ``+99.99'' means that \snrr\,$>$\,100 but that the exact value is not given.

\item column\,14: Note on the cross-identification as described above in the text (Notes). 

\end{itemize}

For those targets that are identified with an object from the KIC, extra information extracted from KIC is given in Table\,\ref{tab:databaseKIC}. There are 17\,241 observed targets for which there are no KIC parameters available.
Apart from the first two columns, which are identical to those given in Table\,\ref{tab:database1D}, Table\,\ref{tab:databaseKIC} contains the following extra columns:

\begin{itemize}

\item column\,3: The KIC right ascension in hh:mm:ss.sss (RA).  

\item column\,4: The KIC declination in dd:mm:ss.ss (DE).

\item column\,5: The angular separation in arcmin between the equatorial coordinates of the fiber (columns\,9 and 10 of Table\,\ref{tab:database1D}) and the equatorial coordinates of the KIC object it was cross-identified with (columns\,3 and 4 of this table) (Ang)

\item column\,6: The contamination factor of objects for which \kepler\ photometry is available (Con). 
The \kepler\ light contamination can be found for different seasons (season 0, 1, 2 and/or 4) with the \kepler\ Target search tool\footnote{Parameters for \kepler\ targets can be found with the \kepler\ Target search tool: {\tt http://archive.stsci.edu/kepler/kepler\_fov/search.php}}, where it is defined as the fraction of light in the aperture that is not due to the target star.
We list the average value of the available seasons. The number of seasons used is given between brackets. (see discussion in Section\,\ref{sect:contamination})

\item column\,7: The \kepler\ \Kp\ magnitude in mag (\Kp).

\item column\,8: The difference between the magnitude as given in headers of the \lamost\ 1D fits file (column\,12 of Table\,\ref{tab:database1D})) and the \kepler\ \Kp\ magnitude of the KIC object it was cross-identified with (column\,5 of this table) (MagDiff). 

\item column\,9: The KIC effective temperature in K (\teff).

\item column\,10: The KIC surface gravity in dex (\logg).

\item column\,11: The KIC metallicity in dex (\mh).    

\item column\,12: The KIC reddening in mag (\ebv).

\item column\,13: The total proper motion in arcsec/yr (pm).

\item column\,14: The angular changes per year in right ascension in arcsec/yr (pmRA). 

\item column\,15: The angular changes per year in right declination in arcsec/yr (pmDE). 

\item column\,16: Note on the cross-identification as described above in the text (Notes). 

\end{itemize} 

\noindent These parameters are not available for every target in the KIC.

Tables\,\ref{tab:database1D} and \ref{tab:databaseKIC} serve as input catalogs for the comparison of the results obtained in the other papers of this series.

\subsection{Examples}
\label{sect:example}

\begin{figure*}
 \begin{center}
 \rotatebox{0}{\includegraphics[height=110mm]{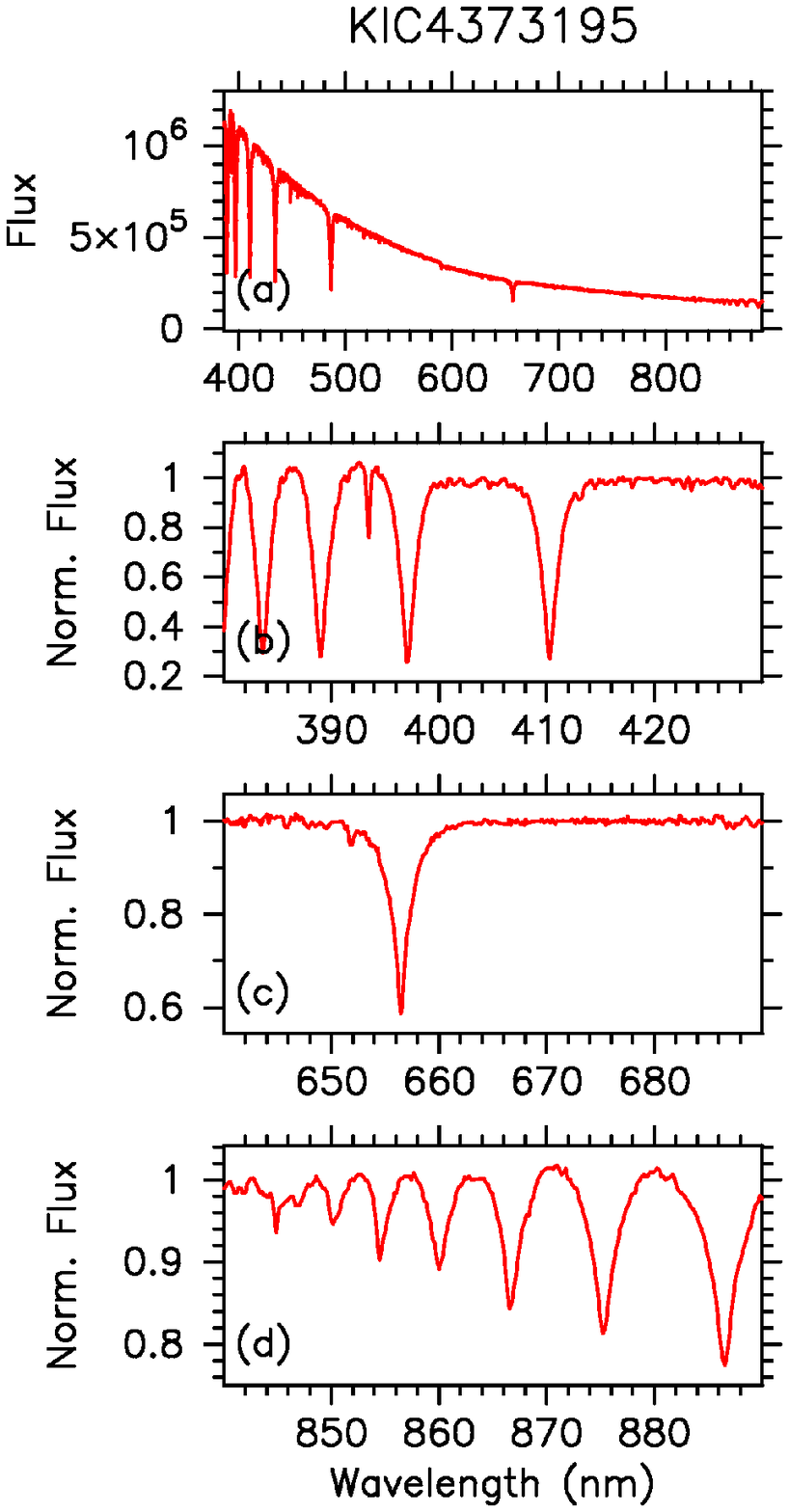}}
 \rotatebox{0}{\includegraphics[height=110mm]{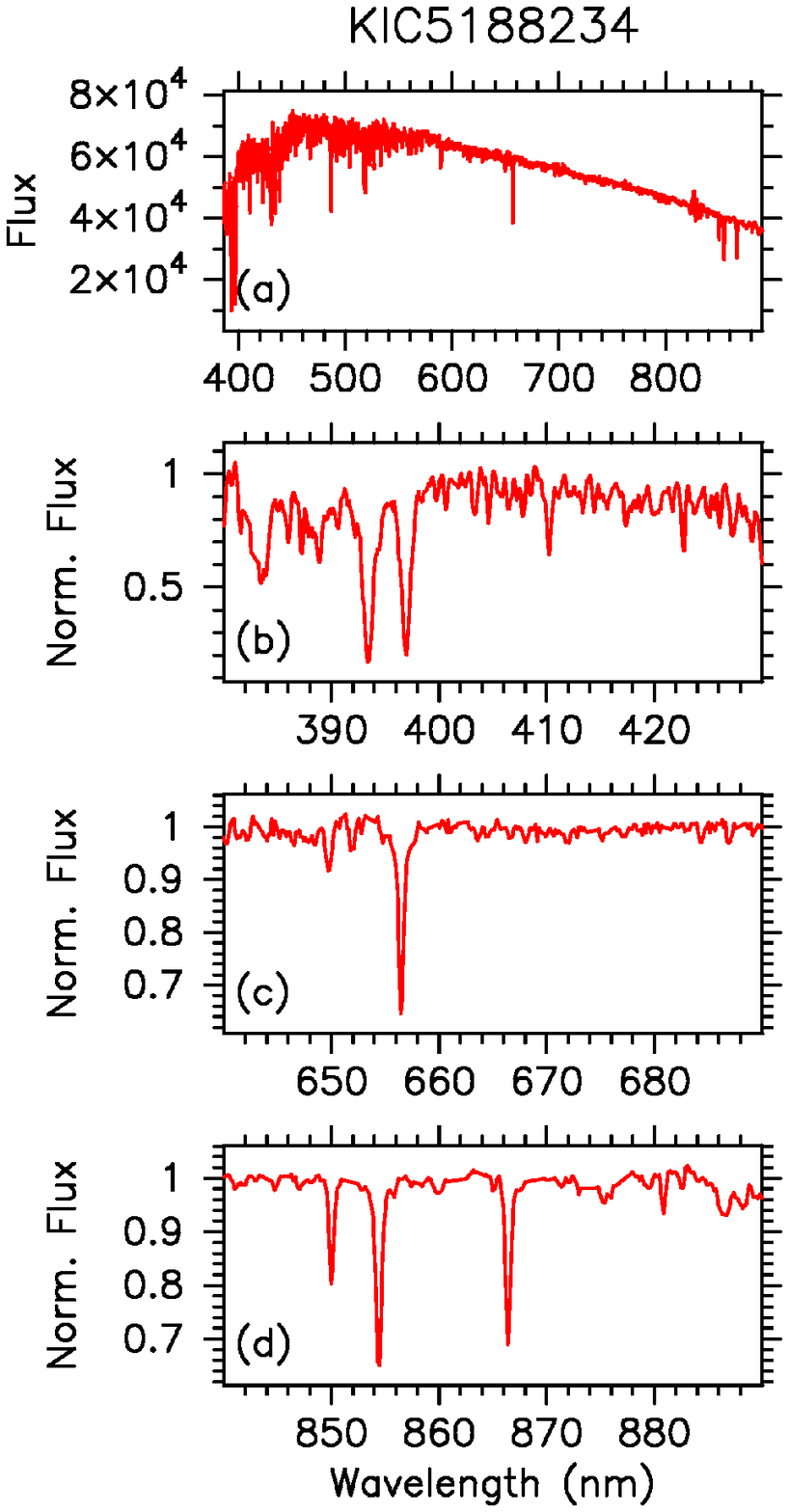}}
 \rotatebox{0}{\includegraphics[height=110mm]{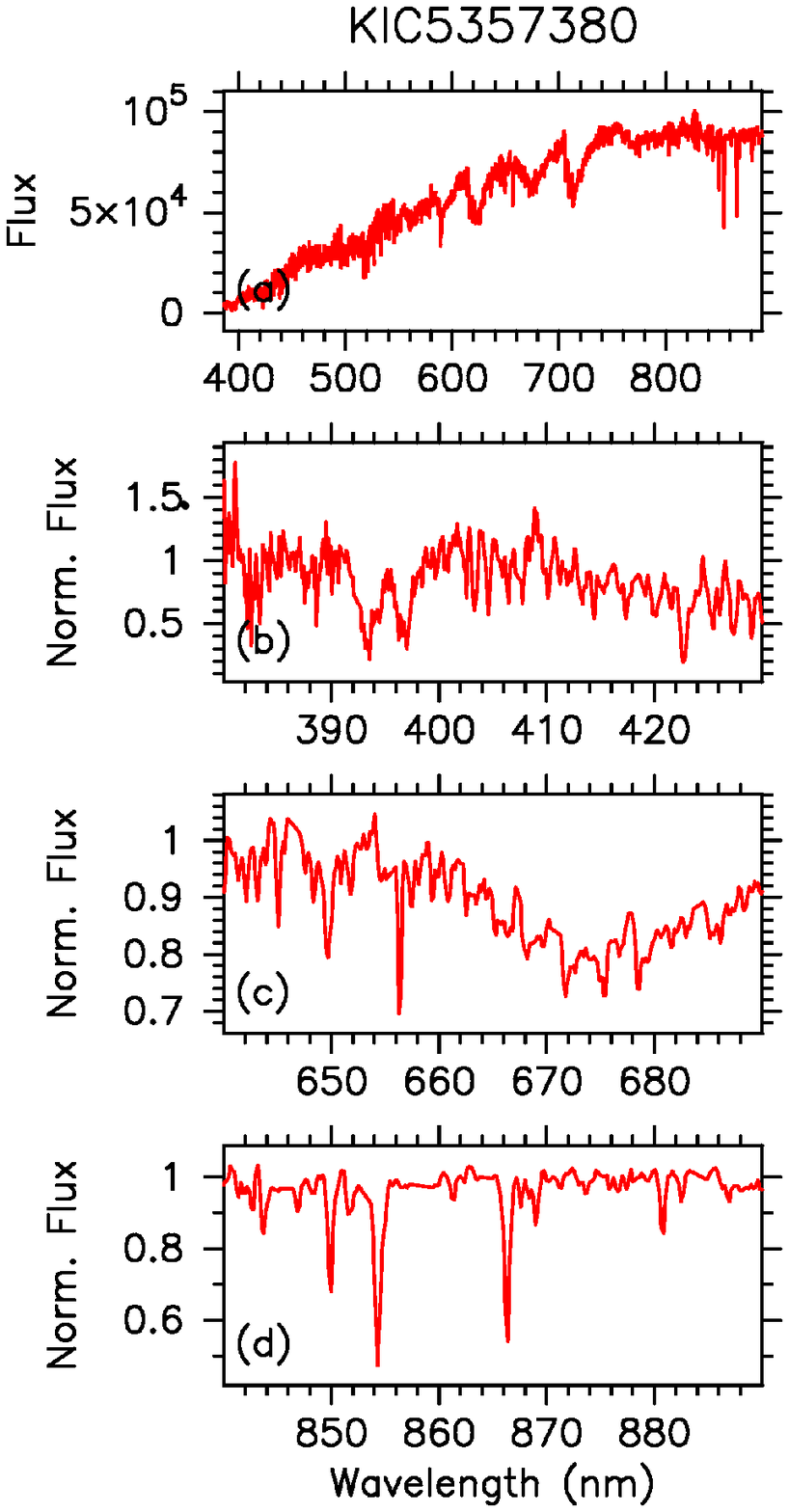}}
 \end{center}
\caption{\label{fig:example} 
Examples of high-quality \lamost\ spectra of KIC\,4373195 (left), KIC\,5188234 (middle) and KIC\,5357380 (right) obtained during the \project. 
For each of these objects we plot the approximately flux calibrated spectra in the full observed wavelength range (panels (a)) and the continuum-normalized fluxes in three different wavelength regions: $380-430$\,nm (panels (b)), $640-690$\,nm (panels (c)) and $840-890$\,nm (panels (d)).
The heliocentric radial velocity corrections were applied during the pipeline reductions. The sampling resolution in pixels varies with wavelength and averages at 2.5 sampling pixels per resolution element \citep{Luo2015arXiv150501570L}.
}
\end{figure*}

\begin{table*}
\begin{center}
\caption{\label{tab:example} 
Representative results obtained by the Asian team (\lasp), the European team (\rotfit) and the American team (\mkclass).
}
  \begin{tabular}{lcclcccl}
    \hline
File Name/Object                        & Version & \snrr  & SpT              & \teff          & \logg        &  \feh         & Method   \\
                                        &         &        &                  & (K)            & (dex)        &  (dex)        &          \\ \hline 
%%===== ~A0-type =====                                                                                                            
%%*** SNR>100 ***                                                                                                                 
spec-56094-kepler05B56094\_2\_sp13-034  & v2.6.4  &  581   &{\it A2\,V       }&  9\,380(56)    & 3.5(11)     & -0.35(5)      & \lasp   \\
KIC\,4373195                            & v2.4.3  &        &{\it B9\,IV      }&  9\,707(589)   & 3.83(11)     & -0.16(14)     & \rotfit  \\
$\ast$                                  & v2.7.5  &        &{\it A0\,IV-V    }&                &              &               & \mkclass \\
%%*** SNR~50  ***                                                                                                                   
spec-56096-kepler08B56096\_1\_sp07-250  & v2.6.4  &   66   &{\it A2\,V       }& 11\,223(113)   & 3.79(18)     & -0.18(8)      & \lasp   \\
KIC\,3137033                            & v2.4.3  &        &{\it B9\,III     }& 10\,839(342)   & 4.00(11)     & -0.08(13)     & \rotfit  \\
                                        & v2.7.5  &        &{\it B9\,V       }&                &              &               & \mkclass \\
%%*** SNR~20  ***                                                                                                                   
spec-56096-kepler08B56096\_2\_sp05-225  & v2.6.4  &   23   & A1\,V          &{\it 10\,869(241)}& 3.87(25)     & -0.41(15)     & \lasp   \\
KIC\,3654218                            & v2.4.3  &        & B9\,IV         &{\it 10\,336(248)}& 3.88(10)     & -0.26(13)     & \rotfit  \\
                                        & v2.7.5  &        & A1\,IV           &                &              &               & \mkclass \\ \hline
%%===== ~F0-type =====                                                                                                              
%%*** SNR>100 ***                                                                                                                   
spec-56096-kepler08B56096\_1\_sp02-050  & v2.6.4  &  122   &{\it A9\,V       }&{\it 7\,168(13)}& 4.16(2)      &{\it -0.42(3)} & \lasp   \\
KIC\,3446837                            & v2.4.3  &        &{\it F0\,V       }&{\it 6\,989(95)}& 4.08(11)     &{\it -0.13(13)}& \rotfit  \\
                                        & v2.7.5  &        &{\it F2\,IV-V\,Fe}-0.6&            &              &               & \mkclass \\
%%*** SNR~50 ***                                                                                                                    
spec-56094-kepler05B56094\_sp16-201     & v2.6.4  &   65   &{\it F5          }&  6\,593(30)    & 4.14(5)      & -0.18(6)      & \lasp   \\
KIC\,5609930                            & v2.4.3  &        &{\it F0\,V       }&  6\,592(215)   & 4.10(13)     & -0.16(15)     & \rotfit  \\
                                        & v2.7.5  &        &{\it F5\,V       }&                &              &               & \mkclass \\
%%*** SNR~20 ***                                                                                                                    
spec-56094-kepler05B56094\_sp12-098     & v2.6.4  &   25   &{\it F5          }&  6\,733(36)    & 4.10(6)      & -0.22(7)      & \lasp   \\
KIC\,5449646                            & v2.4.3  &        &{\it F0\,V       }&  6\,820(281)   & 4.12(11)     & -0.14(14)     & \rotfit  \\
                                        & v2.7.5  &        &{\it F3\,V       }&                &              &               & \mkclass \\ \hline
%%===== ~G0-type =====                                                                                                              
%%*** SNR>100 ***                                                                                                                   
spec-56094-kepler05B56094\_2\_sp09-061  & v2.6.4  &  279   & G0               &{\it 5\,619(12)}&{\it 3.63(4)} &-0.15(3)       & \lasp   \\
KIC\,5188234                            & v2.4.3  &        & G0\,V            &{\it 5\,774(97)}&{\it 4.13(13)}&-0.11(17)      & \rotfit  \\
$\ast$                                  & v2.7.5  &        & G0\,IV-V         &                &              &               & \mkclass \\
%%%*** SNR~50  ***                                                                                                                   
spec-56094-kepler05F56094\_sp15-020     & v2.6.4  &   50   & G0               &  5\,741(41)    & 4.30(12)     & -0.33(10)     & \lasp   \\
KIC\,5095915                            & v2.4.3  &        & G0\,V            &  5\,731(91)    & 4.22(14)     & -0.41(16)     & \rotfit  \\
                                        & v2.7.5  &        & G1\,V            &                &              &               & \mkclass \\
%%*** SNR~20  ***                                                                                                                   
spec-56096-kepler08B56096\_1\_sp05-091  & v2.6.4  &   17   &{\it F9          }&  5\,765(76)    & 3.88(26)     & -0.18(16)     & \lasp   \\
KIC\,3655915                            & v2.4.3  &        &{\it G0\,V       }&  5\,647(216)   & 4.05(25)     & -0.43(20)     & \rotfit  \\
                                        & v2.7.5  &        &{\it G2\,III-IV  }&                &              &               & \mkclass \\ \hline
%%===== ~K0-type =====                                                                                                              
%%*** SNR>100 ***                                                                                                                   
spec-56094-kepler05B56094\_sp07-059     & v2.6.4  &  114   &{\it G5          }&  5\,325(15)    & 4.65(4)      & -0.00(2)      & \lasp   \\
KIC\,2308241                            & v2.4.3  &        &{\it K0\,V       }&  5\,286(141)   & 4.54(15)     &  0.02(18)     & \rotfit  \\
                                        & v2.7.5  &        &{\it K2\,V       }&                &              &               & \mkclass \\
%%*** SNR~50  ***                                                                                                                   
spec-56094-kepler05B56094\_2\_sp15-172  & v2.6.4  &   46   & K1               &  5\,336(28)    & 4.80(8)      &  0.17(5)      & \lasp   \\
KIC\,5442808                            & v2.4.3  &        & K0\,V            &  5\,162(157)   & 4.57(15)     &  0.05(12)     & \rotfit  \\
                                        & v2.7.5  &        & K0\,V            &                &              &               & \mkclass \\
%%*** SNR~20  ***                                                                                                                   
spec-56094-kepler05F56094\_sp13-172     & v2.6.4  &   24   &{\it G5          }&{\it 5\,285(66)}&{\it 4.41(16)}& 0.21(10)      & \lasp   \\
KIC\,4839014                            & v2.4.3  &        &{\it K0\,V      }&{\it 4\,968(191)}&{\it 3.52(40)}& 0.03(14)      & \rotfit  \\
                                        & v2.7.5  &        &{\it K0\,V       }&                &              &               & \mkclass \\ \hline
%%===== ~M0-type =====                                                                                                              
%%** SNR>100 ***                                                                                                                    
spec-56094-kepler05B56094\_2\_sp15-166  & v2.6.4  &  219   & M1               &  3\,717(7)     & 1.49(9)      & -0.05(10)     & \lasp   \\
KIC\,5357380                            & v2.4.3  &        & M1\,III          &  3\,744(85)    & 1.47(15)     & -0.07(13)     & \rotfit  \\
$\ast$                                  & v2.7.5  &        & M0.5\,II-III     &                &              &               & \mkclass \\
%%*** SNR~50  ***                                                                                                                   
spec-56096-kepler08B56096\_2\_sp15-026  & v2.6.4  &   44   & M1               &  3\,754(21)    & 1.45(25)     & -0.31(28)     & \lasp   \\
KIC\,6292615                            & v2.4.3  &        & M1\,III          &  3\,787(96)    & 1.52(17)     & -0.09(13)     & \rotfit  \\
                                        & v2.7.5  &        & M0.5\,II         &                &              &               & \mkclass \\
%%*** SNR~20  ***                                                                                                                   
spec-56096-kepler08B56096\_2\_sp13-010  & v2.6.4  &   21   & M0               &  3\,738(18)    & 1.31(19)     &  0.04(22)     & \lasp   \\
KIC\,5725850                            & v2.4.3  &        & M1\,III          &  3\,761(80)    & 1.50(15)     & -0.05(11)     & \rotfit  \\
                                        & v2.7.5  &        & M2\,III          &                &              &               & \mkclass \\ \hline
  \end{tabular}
  \end{center}
The results derived from the \lamost\ spectra for examples of late B-/early A-, late A-/early F-, late F-/early G-, late G-/early K- and late K-/early M-type stars are given from top to bottom.
For each type we give an example of a \lamost\ spectrum with \snrr\ $>$ 100, \snrr\ $\sim$ 50 and \snrr\ $\sim$ 20 (from top to bottom). 
For each spectrum, we give the name of the analyzed \lamost\ spectrum and the KIC\,number of the observed target (File Name/Object; the \lamost\ spectra indicated with $\ast$ are shown in Fig.\,\ref{fig:example}), the version of the pipelines that was used for each method (Version), the spectral classification (SpT), the effective temperature in K (\teff), the surface gravity in dex (\logg), the metallicity in dex (\feh) and the analysis method (Method; see Section\,\ref{sect:example} for more details about the methods used).
The formal errors are given between brackets in units of the last decimal.
The stellar parameters that don't agree within the formal errors and the spectral sub-classes with a range of more than two are given in italics.
\end{table*}

\begin{figure*}
 \begin{center}
 \includegraphics[width=55mm]{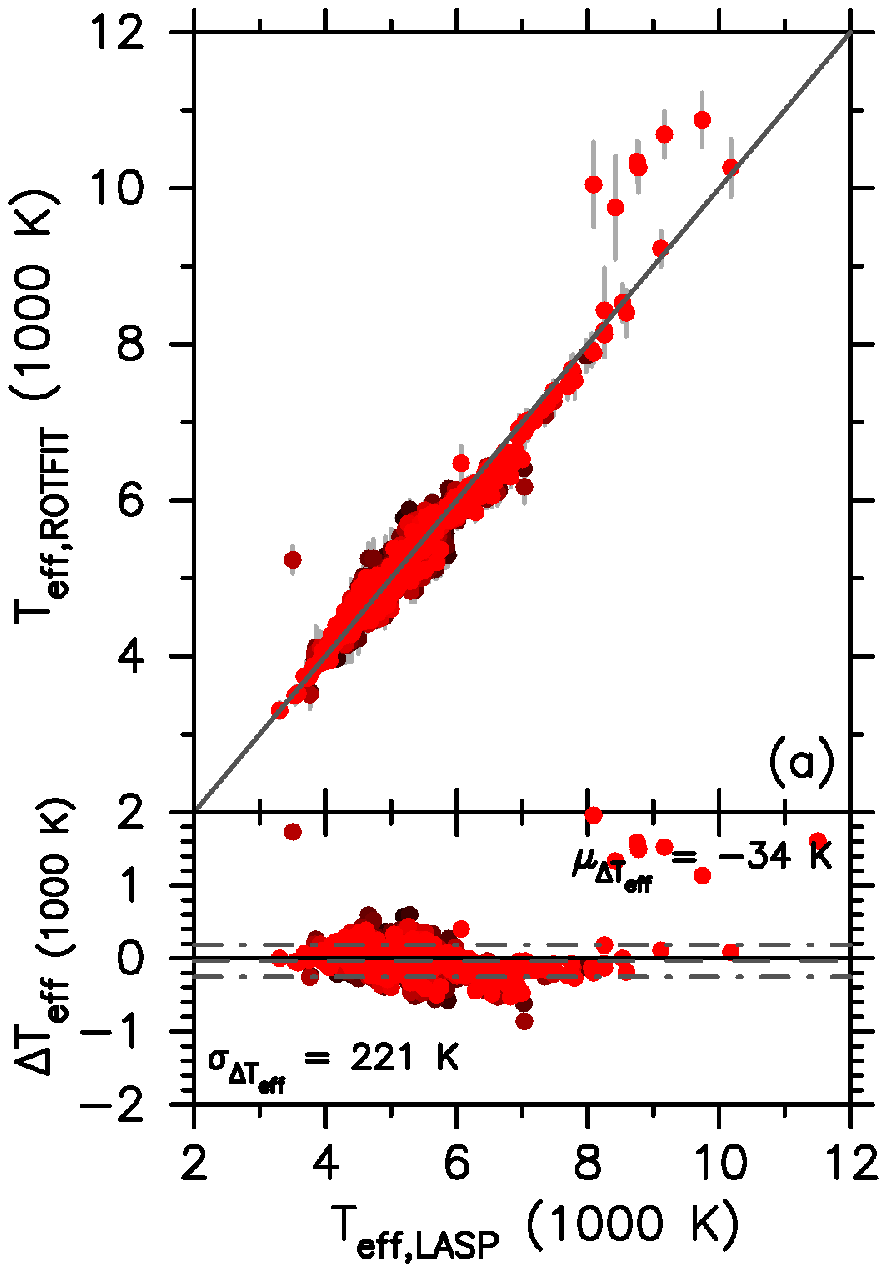}
 \includegraphics[width=55mm]{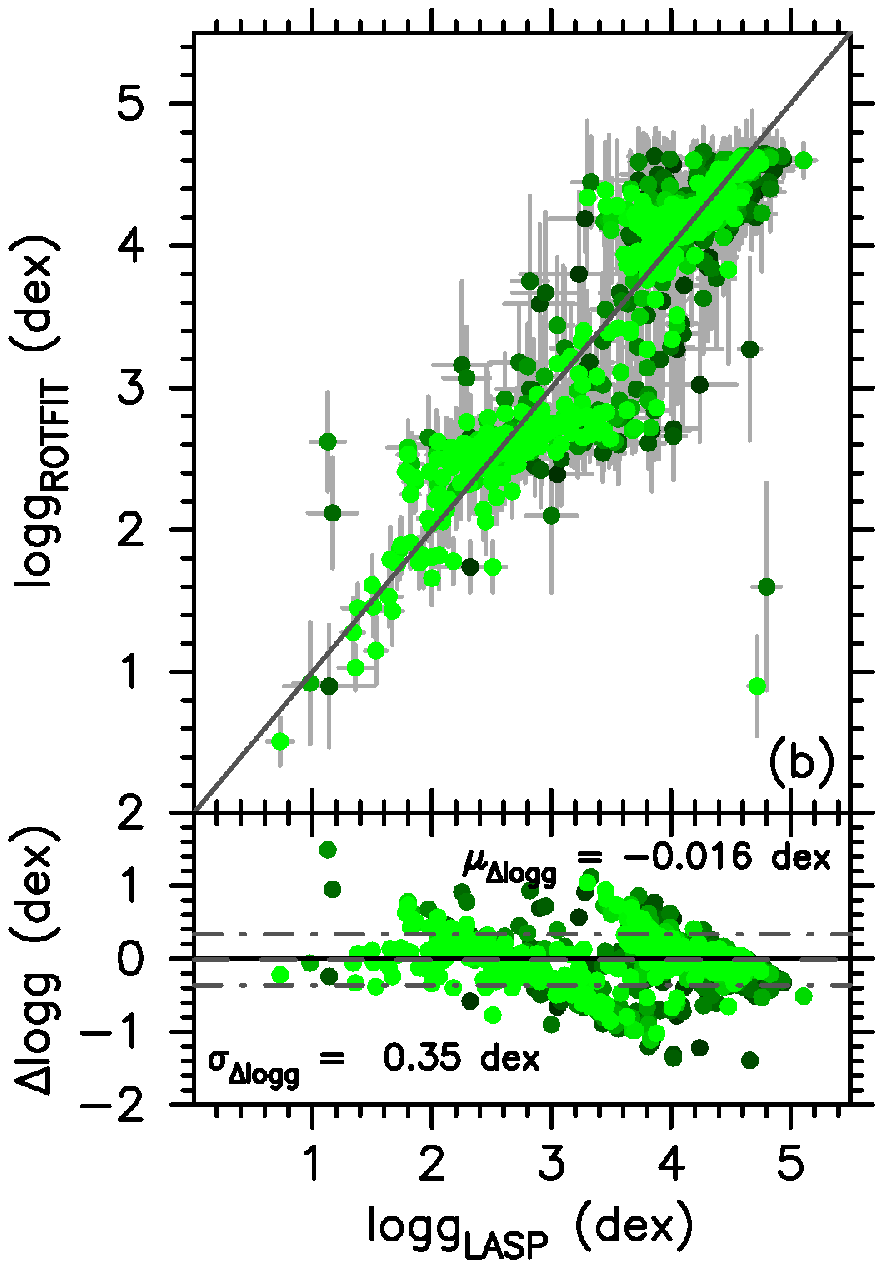}
 \includegraphics[width=55mm]{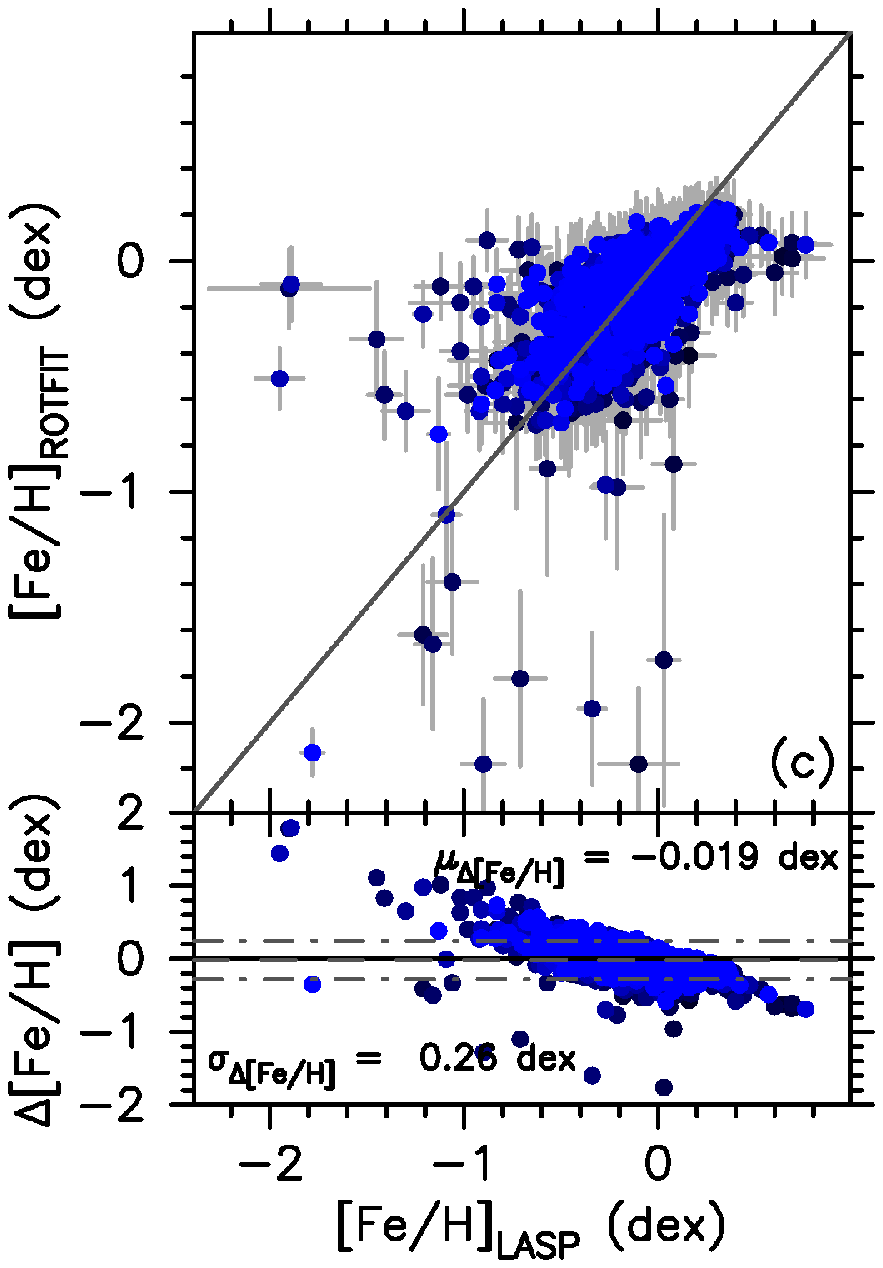}
 \end{center}
\caption{\label{fig:statistics} 
Statistical comparison of the stellar parameters obtained by the Asian team (\lasp) and the European team (\rotfit) from the \lamost\ spectra observed in 2011. The results for \teff, \logg\ and \feh\ are given in red, green, and blue in panels (a), (b), and (c), respectively. The intensity of the color is linked to the \snrr\ value of the \lamost\ spectrum from which the results were derived, going from black for \snrr\,=\,0 to light for \snrr\,$\ge$\,100. In the top panels, we give the derived values as colored dots, the errors as light gray lines and the bisector as a full dark gray line. In the bottom panels, we show the residuals as colored dots, the mean $\mu$ (given in the top right corner) as a dark gray dashed line, and the mean $\pm$ the standard deviation $\sigma$ (given in the bottom left corner) as dark gray dash-dotted lines.
}
\end{figure*}

As an example of the output of \lamost\ reduction pipeline, we show in Fig.\,\ref{fig:example} the spectra of three stars with a very different spectral type. 
The spectra are those of KIC\,3756031 (left panels), KIC\,8235498 (middle panels), and KIC\,5357380 (right panels) classified by the 1D pipeline as A0, G0 and M1 type stars, respectively. 
The top panels (Fig.\,\ref{fig:example} (a)) show the full 1D ``flux-calibrated'' spectra, while the lower panels display three portions of those spectra automatically normalized to the local continuum or pseudo-continuum. 
The most relevant lines, such as the high members of the Balmer series for the hot stars or the Ca\,\textsc{ii}\,H \& K lines for the cool ones, are evident in panels (b) of Fig.\,\ref{fig:example}. 
The other two spectral segments contain H$\alpha$ and the near-IR region (panels (c) and (d) of Fig.\,\ref{fig:example}, respectively). 
For hot stars, the near-IR region clearly shows several Paschen lines while the Ca\,\textsc{ii} infrared triplet lines are visible for cool stars. 
All these spectral features are fundamental for classification purposes. 

There are three teams characterizing all the observed targets by analyzing all the available \lamost\ spectra with a sufficient quality. 
They all have a different approach so the results of the different teams are independent. 

The ``Asian team'' (Ren et al., in preparation) is performing a statistical analysis of the stellar parameters resulting from the \lamost\ stellar parameter pipeline (\lasp; \citealt{Luo2015arXiv150501570L}). It determines the stellar parameters (\teff, \logg, \feh), \vrad\ and the spectral subclass. For the initial guess of the stellar parameters, the Correlation Function Interpolation (\cfi) method is used \citep{Du2012SPIE.8451E..37D}. It searches for the best fit in a grid of 8\,903 synthetic spectra that was build by using the Kurucz spectrum synthesis code based on the {\sc atlas}9 stellar atmosphere models provided by \citet{Castelli2003IAUS..210P.A20C}. The final values are determined with a version of the Universit\'e de Lyon Spectroscopic analysis Software (\ulyss; \citealt{Koleva2009A&A...501.1269K}) adapted to the {\sc lamost} data. This is done by minimizing the squared difference between the observations and models. For the construction of the model corresponding to a set of values of the stellar parameters, an interpolator is used that runs over the whole {\sc elodie} library \citep{Prugniel2001A&A...369.1048P,Prugniel2004astro.ph..9214P,Prugniel2007astro.ph..3658P}. For more details about the methods of the Asian team, we refer to \citet{Wu2011A&A...525A..71W,Wu2011RAA....11..924W}.

The ``European team'' (Frasca et al., in preparation) is determining the stellar parameters and the spectral classification with an adapted version of the code \rotfit\ \citep{Frasca2003A&A...405..149F,Frasca2006A&A...454..301F}.
 The observed spectra are fitted to those available in a grid of spectra for a selection of more than 1000 comparison stars with known stellar parameters from the Indo-U.S. Library of Coud\'e Feed Stellar Spectra \citep{Valdes2004ApJS..152..251V}. Because these spectra have a higher spectral resolution compared to {\sc lamost}, they had to be degraded to match the low-resolution {\sc lamost} spectra. As the library spectra are in the laboratory rest frame and are corrected for their heliocentric \vrad, they also serve as templates to derive \vrad\ with the cross-correlation technique. Moreover, the analysis method is also capable of giving a rough estimation of \vsini\ for rapidly rotating stars. 

The ``American team'' (Gray \& Corbally, in preparation) has developed the code \mkclass\ for classifying stars automatically on the MK spectral classification system independent of the stellar parameter determination \citep{Gray2014AJ....147...80G}. This method requires a library of spectral standards and is designed to classify stellar spectra by direct comparison with MK standards using the same criteria as human classifiers. \mkclass\ is capable of recognizing many of the common spectral peculiarities including barium stars and carbon-rich giants, Ap and Am stars, $\lambda$ Bootis stars, helium-weak and helium-strong B-type stars, etc. For the \lamost\ classifications, the flux-calibrated standards library with 0.36\,nm-resolution spectra obtained using the GM spectrograph at the Dark Sky Observatory of Appalachian State University is employed. Therefore, the \lamost\ spectral resolution is slightly degraded to match that of the spectral library. The accuracy of the resulting classifications does not depend upon the accuracy of the flux calibration of the \lamost\ spectra. Based on tests on spectra classified by humans, the systematic error and standard deviation of the spectral and luminosity classes are 0.1 and 0.5 spectral sub-classes (where a unit spectral subclass is the difference between, for instance, F5 and F6) and 0.02 and 0.5 luminosity classes (where a unit luminosity class is the difference between, for instance, a dwarf (V) and sub-giant (IV) classification), respectively. Thus the accuracy of \mkclass\ is similar to the level of agreement obtained by two independently working skilled human classifiers.

In this paper, the results obtained by the different teams for a few example files are given only to illustrate their compatibility.
Table\,\ref{tab:example} lists the stellar parameters as determined with \lasp, \rotfit\ and \mkclass\ for a variety of spectral types (late B/early A, late A/early F, late F/early G, late G/early K, and late K/early M).
For each of these spectral types, we show results for a spectrum with a \snrr\ of $>$100, $\sim$ 50 and $\sim$20 from top to bottom. 
In general, the results for the stellar parameters \teff, \logg, and \feh\ are fully compatible with each other.
The few that don't agree within the formal errors are given in italics.
They are found both among high and low \snrr\ \lamost\ spectra.
The agreement in the results for the spectral classification is less obvious.
In the cases where the spectral sub-classes show a range of more than two, the results are given in italics.
These differences are partly due to the bad coverage of certain spectral types in the grids of template spectra that were used for the \lasp\ and \rotfit\ calculations.
For this reason, the results of \mkclass\ should be considered as the most trustworthy. 
However, given the discrepancies in standard templates, the agreement in spectral types, especially between \rotfit\ and \mkclass\ is rather good, even for the lowest \snrr.  

In Fig.\,\ref{fig:statistics}, we show a statistical comparison of the stellar parameters obtained by the Asian team (\lasp) and the European team (\rotfit) based on the \lamost\ spectra observed in 2011. The results for \teff, \logg\ and \feh\ derived from the 1\,035 \lamost\ spectra that have been analyzed by both groups are given in panels (a), (b), and (c), respectively. The overall agreement of the results is very good as the mean value and the standard deviation of the residuals amounts to $-34$\,K and 221\,K for \teff, $-0.016$\,dex and 0.35\,dex for \logg, and $-0.019$\,dex and 0.26\,dex for \feh, respectively. The intensity of the colors that are used are linked to the \snrr\ of the \lamost\ spectrum that is analyzed: black corresponds to \snrr\,=\,0 while the lightest colors correspond to \snrr\,$\ge$\,100. Fig.\,\ref{fig:statistics} shows that the most deviating results tend to originate from \lamost\ spectra with a lower \snrr\ value. For \teff, the agreement for the results of both methods is excellent with no apparent systematic behavior, except for the hottest stars. This is likely due to the fact that the grids of model spectra that are used in the analysis methods are sparse in this part of the H-R diagram. For \logg, the values provided by the two analysis codes agree quite well with each other, but the \rotfit\ results appear to be clustered around 2.5\,dex and 4.0\,dex. This behavior is possibly due to the non-uniform density of the templates in the grid used by \rotfit\ and in the {\sc elodie} library used for interpolating in the \lasp\ code. Indeed, these \logg\ values are those typical for main sequence and giant stars, respectively, which are much more frequent in these spectral libraries than stars with an intermediate \logg\ or outside these ranges.  For \feh, the overall comparison is satisfactory but there seems to be a linear trend in the residuals, that could be due to similar contamination problems. Indeed, the metal poor stars, as well as the very metal rich ones, are far less frequent in these libraries.

Note that all the \lamost\ spectra shown in Table\,\ref{tab:example} were gathered during the pilot survey in June 2012 and that Fig.\,\ref{fig:statistics} only shows results derived from the \lamost\ spectra observed in 2011 during the test phase, so the observation and the pipeline procedures were not fully optimized yet. Hence, the discussion given here can be considered as the worst case scenario. The full results obtained by the different teams will be given in the following papers of this series. A more accurate and detailed statistical comparison of the results can only be done after the publication of the full results obtained by the different analysis teams.

\subsection{Characterization}
\label{sect:characterisation}

\begin{figure*}
 \begin{center}
\resizebox{0.95\textwidth}{!}{\includegraphics{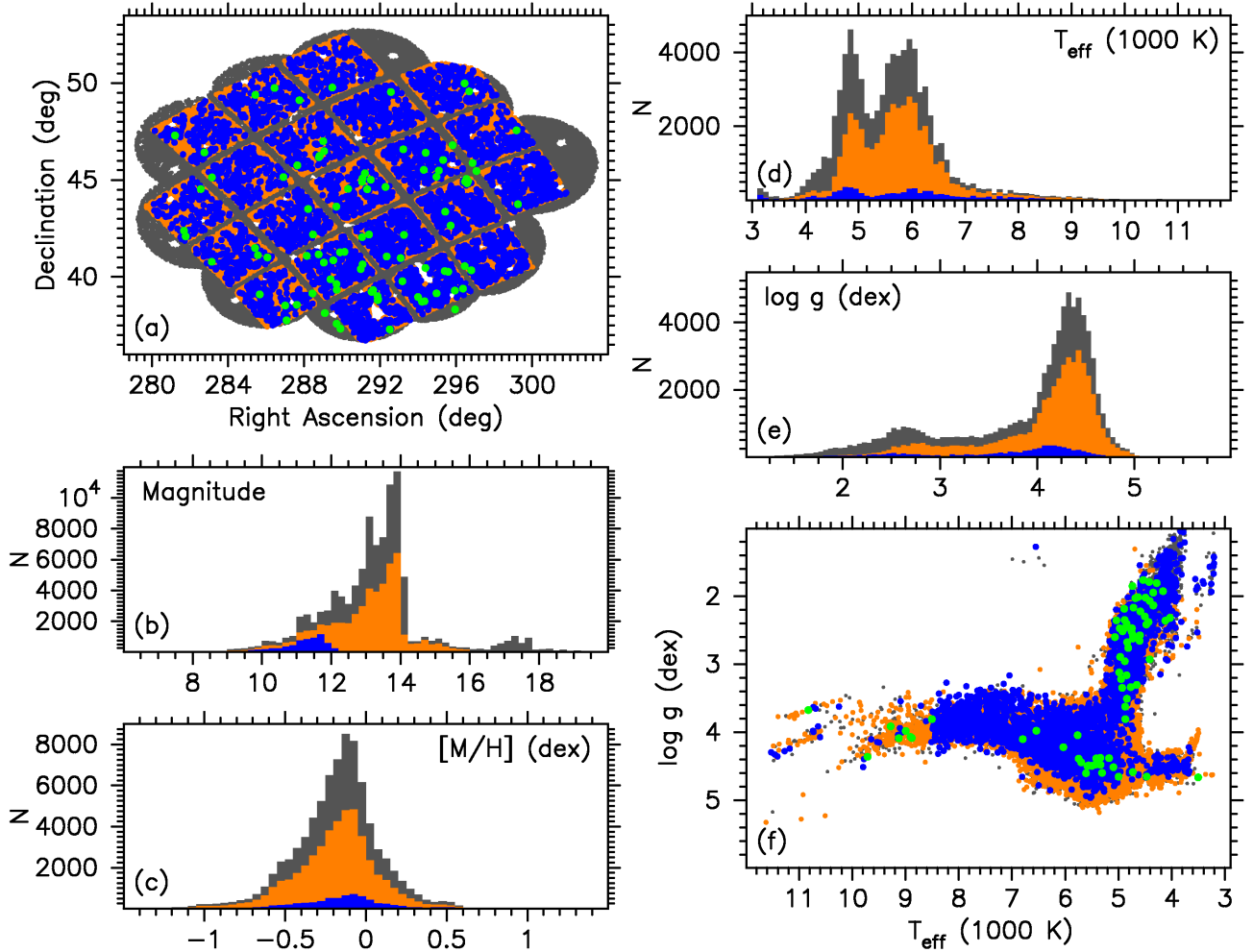}}
 \end{center}
 \caption{\label{fig:all} 
Global characterization of the objects in the database. 
The projection in equatorial coordinates (columns\,9 and 10 of Table\,\ref{tab:database1D}) of the available targets is shown in panel (a). 
We give the distribution of their magnitude (column\,12 of Table\,\ref{tab:database1D}; bin size of 0.2~mag), their KIC metallicity (column\,11 of Table\,\ref{tab:databaseKIC}; bin size of 0.05~dex), their KIC effective temperature (column\,9 of Table\,\ref{tab:databaseKIC}; bin size of 100\,K) and their KIC surface gravity (column\,10 of Table\,\ref{tab:databaseKIC}; bin size of 0.05~dex) in panels (b), (c), (d) and (e), respectively. 
The latter two distributions are represented in an alternative way by showing the position of the objects in a Kiel diagram (\logg\ versus \teff). 
The following color coding is used (from high to low scientific importance; from bottom to top within the distributions): green for standard targets, blue for \kasc\ targets, orange for planet targets, dark gray for extra targets and light gray for field targets. The scientific importance of the different types of targets within the \project\ is also reflected in the size of the symbols in panels (a) and (f). 
(The color representation is only available in the online version of the paper)
}
\end{figure*}

In this section, we discuss a few figures to characterize the database of the \project.

\subsubsection{Objects}
\label{sect:objects}

In Fig.\,\ref{fig:all}, we give an overview of the objects in our database. The different types of targets are indicated with different colors. We use green for standard targets, blue for \kasc\ targets, orange for planet targets, dark gray for extra targets and light gray for field targets. The scientific importance of the different types of targets within the \project\ is also reflected in the size of the symbols. 
Note that only a few standard targets and field targets were observed, so they are invisible on some of the panels in Fig.\,\ref{fig:all}. 

The projection in equatorial coordinates is shown in Fig.\,\ref{fig:all} (a). 
From this figure it can be easily seen that the \kepler\ FoV is well covered by observations with the \lamost.
Only few parts are not covered: the center and off-center holes of the \pointings\ (containing the central bright star and the guide stars, respectively; cf. Sect.\,\ref{sect:pointings}) and the edge of one \kepler\ CCD (near right ascension $\sim$296\degree\ and declination $\sim$51\degree).
It also shows that there is overlap between the observed \pointings.

Fig.\,\ref{fig:all} (b) shows the magnitude distribution of the objects (column\,12 of Table\,\ref{tab:database1D}). 
The largest concentration of the targets is found between magnitudes 12.5 and 14 which is a reflection of our observation strategy. 
Indeed, the observations only focused on very bright objects (\Kp\,$<$ 10.5) on two occasions (plates 110608\_1 and 120604\_1 in Table\,\ref{tab:LKspectra}) and on faint objects on four occasions (\Kp\,$>$ 14.0 for plates 120615\_3 and 120617\_3, and 140520\_1, and \Kp\,$>$ 16.8 for plate 131002\_1 in Table\,\ref{tab:LKspectra}).

Fig.\,\ref{fig:all} (c), (d) and (e) give the distributions of the KIC stellar parameters of the observed objects: \mh\ (column\,11 of Table\,\ref{tab:databaseKIC}), \teff\ (column\,6 of Table\,\ref{tab:databaseKIC}) and \logg\ (column\,10 of Table\,\ref{tab:databaseKIC}), respectively. 
The latter two distributions are visualized in a different way in Fig.\,\ref{fig:all} (f) where the objects are placed in a Kiel diagram (\logg\ versus \teff), which is an alternative version of the Hertzsprung-Russell (H-R) diagram. 
The distribution of the KIC \mh\ suggests that the majority of the objects are slightly metal-poor with a median value around -0.4. 
A subset of 12\,000 \lamost\ spectra of stars from the \project\ have been used by \citet{Dong2014ApJ...789L...3D} to study the reliability of the metallicity values as given in the KIC catalog. 
They found that the true metallicity and the dynamic range of the \kepler\ sample are both systematically underestimated by the KIC.
Hence, the KIC \mh\ values should be treated with caution.
The majority of our objects have a KIC \teff\ between 4\,000 and 7\,000\,K. 
The \kepler\ field contains only few hot stars for which the parameter determination based on the Sloan photometry suffers the most from the lack of an ultraviolet filter. The double peak structure of the \teff\ and \logg\ distribution reflects the superposition of the main sequence (centered around \teff\ $\sim$6\,000\,K and \logg\ $\sim$4.5\,dex) and the red giant branch (centered around \teff\ $\sim$4\,800\,K and \logg\ $\sim$2.7\,dex).

There are no significant differences between the distributions of the different groups of targets, except for the spatial and brightness distribution: the standard targets are very bright, the \kasc\ targets are mainly restricted to magnitude 12 while the other types of target extend to fainter targets. Obviously, the gaps between the \kepler\ CCDs are filled by extra (dark gray dots) and field (light gray) targets (Fig.\,\ref{fig:all}a). 
The currently available \lamost\ spectra of each type of target in itself (except the standard and field targets that are few) are appropriate to study their stellar parameters, thus tracing the target properties across the H-R diagram.

\subsubsection{Multiple observations}
\label{sect:multiple}

As there is considerable overlap between some of the observed \pointings\ (Fig.\,\ref{fig:LKfields} and Fig.\,\ref{fig:all} (a)), there are many objects that have been observed more than once: 14\,186 objects have been observed two times, 2\,483 objects three times, 332 objects four times and 113 objects at least five times (cf. bottom rows of Table\,\ref{tab:LKfields_ok}).
Many of these objects are indeed found in the overlapping regions. Those within the non-overlapping regions are either objects that are observed on different plates of the same field or objects that are observed as a field star that turned out to be an already observed \kepler\ star (or vice versa). 
Even though the main aim of our \project\ is to determine the stellar parameters of as many objects within the \kepler\ FoV as possible, it is extremely valuable to have a significant number of the objects with multiple \lamost\ spectra as they can be used to check the consistency of the results of the analysis methods used and to discover stars with a variable \vrad.
The distributions of the parameters for the sub-sample of stars with multiple exposures are consistent with the general distributions shown in Fig.\,\ref{fig:all}. Hence, these targets can be used to check the consistency of the results across the H-R diagram.

\subsubsection{Signal-to-noise ratio}
\label{sect:snr}

\begin{figure}
 \begin{center}
  \resizebox{0.45\textwidth}{!}{\includegraphics{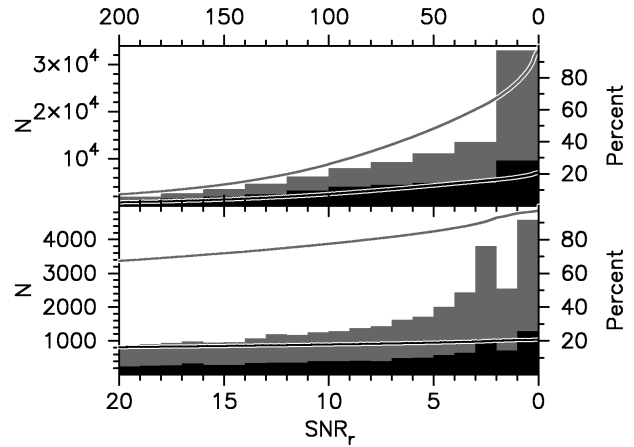}}
 \end{center}
 \caption{\label{fig:snr} 
Histogram of the SNR-values in the r-band (\snrr) of the \lamost\ spectra (cf. column\,13 of Table\,\ref{tab:database1D} for the regions [200,0] (bin size 20; top) and [20,0] (bin size 1; bottom panel).
The colors gray and black refer to the full data set (101\,086 spectra) and to the full data set restricted to the spectrum with the highest \snrr\ for the targets that have been observed by the \kepler\ mission (42\,209 spectra), respectively.
The full lines give the cumulative percentage of \lamost\ spectra for each data set as a function of decreasing \snrr\ value.
}
\end{figure}

Fig.\,\ref{fig:snr} gives the distribution of the SNR in the r-band (\snrr) of the available \lamost\ spectra (column\,13 of Table\,\ref{tab:database1D}).
It gives an overview of the overall quality of the data.
On this figure, we give a histogram of the \snrr\ values for two data sets: the full set of 101\,086 \lamost\ spectra that have been reduced (in gray) and the full data set restricted to the 42\,209 \lamost\ spectra with the highest \snrr\ for the targets that have been observed by the \kepler\ mission (in black).
The top and bottom panels give a global overview for \snrr\ up to 200 and a detailed overview for the \snrr\ range $[0,20]$, respectively.
The small fraction of \lamost\ spectra with \snrr\,$\ge$\,200 is not shown.
We consider \snrr\,= 10 as the strict minimum requirement for a \lamost\ spectrum to be analyzed and \snrr\,= 20 as the minimum requirement for an object to be removed from our target list for future observations.
The full lines plotted on top of the histograms in each panel of Fig.\,\ref{fig:snr} give the cumulative percentage of \lamost\ spectra for each data-set as a function of decreasing \snrr\ value. 
About 67.3\,percent of the reduced \lamost\ spectra have a \snrr\,$\ge$\,20 and about 77.5\,percent have a \snrr\,$\ge$\,10.
So far, we have \lamost\ data for 21.1\,percent of the 199\,718 objects that have been observed during the \kepler\ mission, but not all of these \lamost\ spectra are of a sufficient quality.
For 17.9\,percent of these \kepler\ mission targets, we have a \lamost\ spectrum with a \snrr\ above 10.
For 16.3\,percent of them, the \snrr\ of their best \lamost\ spectrum exceeds 20.
These latter targets can be removed from the target list for future \lamost\ observations.

\begin{figure}
 \begin{center}
  \resizebox{0.45\textwidth}{!}{\includegraphics{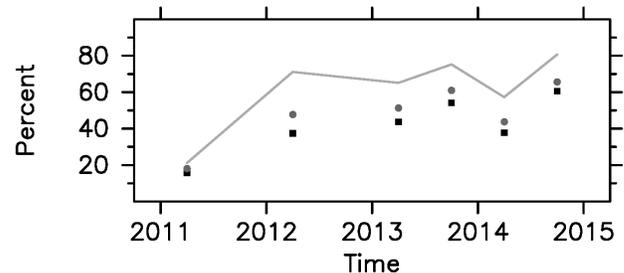}}
 \end{center}
 \caption{\label{fig:snrfiber} 
Overview of the evolution of the success rate of the \lamost\ during the \project\ in time.
We show the percentage of the fibers leading to a successfully reduced \lamost\ spectrum (light gray full line) and to a \lamost\ spectrum with \snrr\,$\ge$\,10 (gray circle) and \snrr\,$\ge$\,20 (black square).
The observations are subdivided into semesters: May-June and September-October of each year.
}
\end{figure}

The \snrr\ can also be used to estimate the percentage of \lamost\ fibers leading to high quality data (\snrr\,$\ge$\,20) as a function of time.
This is illustrated in Fig.\,\ref{fig:snrfiber}.
During the test phase, high quality \lamost\ spectra emerged from about 15\,percent of the fibers only while this percentage reached about 60\,percent during the last semester.
This is mainly due to an improvement in the fiber positioning and the better calibration of the \lamost\ spectra. Also the meteorological circumstances during the observations can have an effect on the overall quality of the \lamost\ spectra. They can cause seasonal fluctuations but can not be responsible for the clear upwards trend as seen in Fig.\,\ref{fig:snrfiber}.

%        #spectra  SNR20   SNR10  #plates #fibers    %spectra      %SNR20  %SN10
%2011.1     #2543  74.83%  85.53%    3    #12000 -->   21.19% -->  15.85%  18.12%
%2012.1    #19903  52.63%  67.19%    7    #28000 -->   71.08% -->  37.41%  47.76%
%2013.1     #7817  67.07%  78.81%    3    #12000 -->   65.14% -->  43.69%  51.34%
%2013.2    #33101  71.96%  81.09%   11    #44000 -->   75.23% -->  54.13%  61.00%
%2014.1    #18341  65.93%  76.29%    8    #32000 -->   57.32% -->  37.79%  43.73%
%2014.2    #19381  75.06%  81.30%    6    #32000 -->   80.75% -->  60.61%  65.65%

%2013      #40918  71.03%  80.65%   14    #56000 -->   73.07% -->  51.90%  
%2014      #37722  70.62%  78.86%   14    #56000 -->   67.36% -->  47.56%  
%all      #101086  67.34%  77.45%   38   #152000 -->   66.50% -->  44.78%  51.50% 

\subsubsection{Contamination}
\label{sect:contamination}

\begin{figure}
 \begin{center}
  \resizebox{0.49\textwidth}{!}{\includegraphics{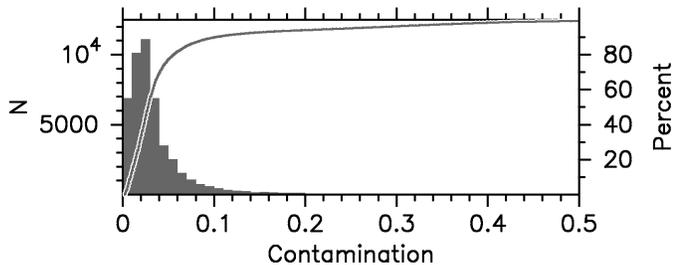}}
 \end{center}
 \caption{\label{fig:cont} 
Overview of the \kepler\ contamination factors for objects in the database (column\,6 of Table\,\ref{tab:databaseKIC}). 
A histogram with a bin size of 0.01 is given.
The full lines give the cumulative percentage of targets as a function of increasing contamination factor.
}
\end{figure}

The \kepler\ contamination factor of an object tells us how much the \kepler\ light curves are affected by the presence of neighboring objects. The higher the \kepler\ contamination factor, the more overlap with neighboring objects and hence the less reliable the results that can be derived from the \kepler\ data. 
As the \lamost\ fiber size and typical seeing match the \kepler\ pixel size, the \kepler\ contamination factor can be used as an indication of how much the ground-based measurements of \lamost\ can be expected to be affected by contamination effects.
The \kepler\ contamination factor was not taken into account for the construction of the prioritized list of targets (see Section\,\ref{sect:target}). 
In Fig.\,\ref{fig:cont}, we give the distribution of the \kepler\ contamination factor for the 50\,053 observed objects for which a value can be found with the \kepler\ Target search tool. 
We used a bin size of 0.01 for the construction of the histogram. 
The full lines give the cumulative percentage of targets as a function of increasing contamination factor.
The distribution peaks around a \kepler\ contamination factor of 0.025 and 89.77\,percent of the objects have a value below 0.1, which means that the contamination effects for the vast majority of the objects are small. 
For 365 objects (0.73\,percent), the \kepler\ contamination factor is higher than 0.5 which means that more than 50\,percent of the counts in the light curves of these objects comes from neighboring objects. 
The \lamost\ data for such objects should be interpreted with caution. 
We cannot rule out that in some cases that the wrong star was observed.

%%%%%%%%%%%%%%%%%%%%%%%%%%%%%%%%%%%%%%%%%%%%%%%%%%%%%%%%%%%%%%%%%%%%%%%%%%%%%%
\section{Discussion}
\label{sect:discussion}
%%%%%%%%%%%%%%%%%%%%%%%%%%%%%%%%%%%%%%%%%%%%%%%%%%%%%%%%%%%%%%%%%%%%%%%%%%%%%%

The \lamost\ spectra that are described in the database of this paper have been proven to be of a high scientific value, especially for the large community involved in the \kepler\ research. 
The \project\ has led to the first spectroscopic observation for many, in general relatively faint, \kepler\ targets. 
To maximize the visibility of the \lamost\ spectra within this community, the observations are listed in the overview table with ground-based follow-up observations for \kepler\ targets on the web page of the \kepler\ Asteroseismic Science Operations Center ({\sc kasoc})\footnote{The {\sc kasoc} web page ({\tt http://kasoc.phys.au.dk/}) is a private web page that can be accessed by registered \kasc\ members only.}.

The \lamost\ spectra have already been used by several groups of astronomers.
Apart from the three teams that are deriving stellar parameters from all the good-quality \lamost\ spectra, the data of the \project\ has been useful to select interesting targets for other types of studies.
The wavelength coverage of the \lamost\ observations include the Lithium line at $\sim$670\,nm, making this survey ideal for identifying candidates with unusual Li abundances. 
Of particular interest are those \kasc\ targets where oscillations have been detected, and allow to put stringent constraints on the evolutionary phase and physical properties of the star. 
Thanks to our \lamost\ survey, \citet{SilvaAguirre2014ApJ...784L..16S} have detected the first confirmed Li-rich core-helium-burning giant.
The \lamost\ observations also include the Ca\,\textsc{ii}\,H \& K lines at 396.85 and 393.37\,nm, respectively (see Fig.\,\ref{fig:example} (b)), making it possible to measure the chromospheric activity of the observed stars. 
The most commonly used expression for stellar chromospheric activity is the dimensionless $S$ index \citep{Baliunas1995ApJ...438..269B} and the excess flux \citep{Hall2007AJ....133.2206H}. 
Karoff et al. (in preparation) have measured both for 7\,700 main-sequence solar-like stars with \teff\ between 5\,000 and 6\,000 K in the \kepler\ field with \lamost\ spectra using the same method that was used by \citet{Karoff2013MNRAS.433.3227K}.
Bostanc{\i} et al. (submitted to MNRAS) used the \lamost\ spectra of members of NGC\,6866 to derive the metallicity and the \vrad\ of this open cluster.
Some individual \lamost\ spectra of the \project\ have also been used in case studies: e.g. the red giant star KIC\,5689820 \citep{Deheuvels2014A&A...564A..27D}.

Even though the \lamost\ spectra obtained during the test phase and the pilot survey for the \project\ have been proven to be of a sufficient quality for several scientific applications, the \lamost\ observations and procedures are still being subjected to continuous improvements. 
For the general survey, the observation plan is better and resulted in a rapid increase in the data quantity, although more bright stars ($<$ 14 mag) are observed on bright nights.
Also, the quality control has been improved, which makes the success rate of the observations reach the 60\,percent level.
As more of the acquired spectra reach a \snrr\ level of 10, more science data can be collected.
There have already been several updates in the \lamost\ reduction and analysis pipelines.
For the faint objects, the sky subtraction now reaches the SDSS level while there are still some sky continuum problems for bright stars on bright moon nights.
This issue can be solved by a better sky model.
The difficulty of looking for standard stars in some spectrographs has been solved by using the estimates of the stellar parameters obtained before the flux calibration.
Based on these parameter estimates, late F-type stars are selected as standard stars and the flux calibration for one spectrograph is done by using the curve obtained by comparing the observed spectra with model spectra of standard stars.
This makes it possible to connect almost all the blue and red arm spectra and to improve the relative flux calibration.
A more detailed description of the recent improvements and of the data process itself are given in \citet{Luo2015arXiv150501570L}.
Astronomers who need help to process the \lamost\ spectra they are interested in can contact the members of the \lamost\ consortium\footnote{Send requests to Dr. Ali Luo (lal@bao.ac.cn)}.
The prospect for \lamost\ is bright, and the observations for the \project\ will continue.

%%%%%%%%%%%%%%%%%%%%%%%%%%%%%%%%%%%%%%%%%%%%%%%%%%%%%%%%%%%%%%%%%%%%%%%%%%%%%%
\section{Acknowledgments}
%%%%%%%%%%%%%%%%%%%%%%%%%%%%%%%%%%%%%%%%%%%%%%%%%%%%%%%%%%%%%%%%%%%%%%%%%%%%%%

We are thankful for the useful comments and suggestions of the anonymous referee that improved the quality of our paper.
Guoshoujing Telescope (the Large Sky Area Multi-Object Fiber Spectroscopic Telescope \lamost) is a National Major Scientific Project built by the Chinese Academy of Sciences. Funding for the project has been provided by the National Development and Reform Commission. \lamost\ is operated and managed by the National Astronomical Observatories, Chinese Academy of Sciences.
The research is supported by the ASTERISK project (ASTERoseismic Investigations with SONG and \kepler) funded by the European Research Council (Grant agreement no.: 267864), the `Lend\"ulet-2009 Young Researchers' Program, ESA PECS Contract No. 4000110889/14/NL/NDe, and the National Science Foundation under Grant No. NSF PHY05-51164.
Part of the research leading to these results has received funding from the European Research Council under the European Community's Seventh Framework Programme (FP7 2007–2013) under grant agreements No. 227224 (ERC/PROSPERITY), No. 269194 (ASK), and No. 312844 (SPACEINN).
Funding for the Stellar Astrophysics Center (Aarhus, Denmark) is provided by The Danish National Research Foundation (Grant agreement no.: DNRF106). 
JNF, ABR and XHY acknowledge the support from the Joint Fund of Astronomy of National Natural Science Foundation of China (NSFC) and Chinese Academy of Sciences through the Grant U1231202, and the support from the National Basic Research Program of China (973 Program 2014CB845700 and 2013CB834900).
MB is F.R.S.-FNRS Postdoctoral Researcher, Belgium.
SD is supported by ``the Strategic Priority Research Program- The Emergence of Cosmological Structures'' of the Chinese Academy of Sciences (grant No. XDB09000000).
JM\.Z acknowledges the Polish MNiSW grant NN203 405139 and the European Community's Seventh Framework Programme (FP7/2007-2013) ASK grant no. 269194.
KU acknowledges financial support by the Spanish National Plan of R\&D for 2010, project AYA2010-17803.
EN acknowledges support from the NCN grant No. 2014/13/B/ST9/00902.
RSz acknowledges the University of Sydney IRCA grant and was supported by the Hungarian Scientific Researc Fund -- OTKA K-115709.
We thank N. Batalha and D.W. Latham for providing the initial lists of objects that were used to select the planet and extra targets, respectively.

\end{document}